\documentclass[apj]{emulateapj}
\usepackage{lscape}

\newcommand{\co}{\mbox{\rm CO}}

\newcommand{\hi}{\mbox{\rm \ion{H}{1}}}

\newcommand{\kmpers}{\mbox{km~s$^{-1}$}}
\newcommand{\Kkmpers}{\mbox{K~km~s$^{-1}$}}

\newcommand{\xcounits}{\mbox{cm$^{-2}$ (K km s$^{-1}$)$^{-1}$}}

\newcommand{\xco}{\mbox{\rm X$_{\rm CO}$}}

\shorttitle{The Molecular ISM of Dwarf Galaxies}
\shortauthors{Leroy et al.}
\slugcomment{Accepted for publication in \emph{The Astrophysical Journal}}

\begin{document}
% DO THIS LATER
\title{The Molecular ISM of Dwarf Galaxies on Kiloparsec Scales: A New
Survey for CO in Northern, IRAS-detected Dwarf Galaxies}

\author{A. Leroy, A. D. Bolatto, J. D. Simon, L. Blitz}

\affil{Department of Astronomy, 601 Campbell Hall, University of 
       California at Berkeley, CA  94720}
\email{aleroy@astro.berkeley.edu}
\email{bolatto@astro.berkeley.edu}
\email{jsimon@astro.berkeley.edu}
\email{blitz@astro.berkeley.edu}

\begin{abstract}
We present a new survey for CO in dwarf galaxies using the ARO Kitt
Peak 12m telescope. This survey consists of observations of the
central regions of $121$ northern dwarfs with IRAS detections and no
known CO emission. We detect CO in $28$ of these galaxies and
marginally detect another $16$, increasing by about 50\% the number of
such galaxies known to have significant CO emission. The galaxies we
detect are comparable in stellar and dynamical mass to the Large
Magellanic Cloud, although somewhat brighter in CO and fainter in the
FIR. Within dwarfs, we find that the CO luminosity, $L_{CO}$, is most
strongly correlated with the $K$-band and the far infrared
luminosities. There are also strong correlations with the radio
continuum and $B$-band luminosities, and linear diameter. Conversely,
we find that FIR dust temperature is a poor predictor of CO emission
within the dwarfs alone, though a good predictor of normalized \co\
content among a larger sample of galaxies. We suggest that $L_{CO}$
and $L_K$ correlate well because the stellar component of a galaxy
dominates the midplane gravitational field and thus sets the pressure
and density of the atomic gas, which control the formation of H$_2$
from \hi.  We compare our sample with more massive galaxies and find
that dwarfs and large galaxies obey the same relationship between \co\
and the 1.4 GHz radio continuum (RC) surface brightness. This
relationship is well described by a Schmidt Law with $\Sigma_{RC}
\propto \Sigma_{CO}^{1.3}$. Therefore, dwarf galaxies and large
spirals exhibit the same relationship between molecular gas and star
formation rate (SFR). We find that this result is robust to moderate
changes in the RC-to-SFR and CO-to-H$_2$ conversion factors. Our data
appear to be inconsistent with large (order of magnitude) variations
in the CO-to-H$_2$ conversion factor in the star forming molecular
gas.
\end{abstract}

% INTRODUCTION
\section{Introduction}

Although many dwarf galaxies are actively forming stars, detecting
molecular gas in these objects has frequently proven difficult. \co,
the brightest and most abundant tracer of molecular hydrogen, is
generally not seen in dwarfs \citep[see][ and references
therein]{TKS98}. Since stars form out of molecular gas, this lack of
\co\ emission is puzzling. Does the faintness of \co\ reflect genuine
scarcity of molecular hydrogen (H$_2$), or does \co\ become a poor
tracer of H$_2$ in the low metallicity interstellar medium (ISM) of
dwarf galaxies? Several authors have argued for the latter
interpretation based on measurements of the virial and dust masses of
giant molecular clouds (GMCs) \citep[][]{WI95,IS97}.  However, recent
high resolution studies of GMC virial masses in nearby galaxies have
found little evidence for changes in the \co-to-H$_2$ ratio between
dwarf and large galaxies or as a function of metallicity
\citep[][]{WA01, WA02, BOL03, RO03}. These measurements suggest that
CO remains a good tracer of dense molecular gas down to metallicities
of $Z\sim1/4~Z_\odot$.

Molecular gas in dwarf galaxies is of particular interest because
these systems are characterized by lower metallicities, stronger
radiation fields, and shallower potential wells than large
star-forming galaxies. These conditions resemble those in the early
universe and may affect the properties of GMCs in these objects.
Indeed, the GMC mass spectrum appears to vary among Local Group
galaxies \citep[][]{SO87,EN03,HCS01,MIZ01} and some data suggest that
several nearby galaxies contain clouds that obey a different
size-linewidth-luminosity relation than the Milky Way \citep[][ though
differences in resolution leave this matter still
open]{LO98,RA99,WA02}. Are such variations common? If so, what is
their impact on galaxy evolution? GMC properties may affect a cloud's
star formation rate (SFR) or the initial mass function (IMF). In this
case, we would expect galaxies with systematically different GMC
populations to show different relationships between SF tracers and the
molecular gas.

As a step towards resolving these questions, we have carried out a
large \co\ survey of 121 nearby, infrared-bright dwarf galaxies. We
detect or marginally detect $\sim 1/3$ of these galaxies, a higher
success rate than was achieved by previous surveys \citep[e.g.,
][]{ITB95}. Thus, we find that although \co\ is faint in dwarf
galaxies, it is detected when observations reach sufficient
sensitivity. Combining these data with results from the literature
yields a sample of $\sim 80$ nearby dwarf galaxies with known
molecular emission. This large sample of \co-emitting dwarfs allows us
to investigate how the amount of molecular gas relates to other galaxy
properties. We determine the H$_2$-to-SFR relation for this sample and
compare it to that found in large galaxies. We also use the sample in
conjunction with our nondetections to examine which properties of a
galaxy are the best predictors of \co\ emission.

The remainder of this paper is organized as follows. In \S 2, we
describe our survey, including our method for classifying galaxies as
detections or nondetections. In \S 3, we examine the results of our
survey in some detail and discuss what galaxy properties are most
closely related to \co\ emission. In \S 4, we study the relationship
between \co\ and SFR on $\sim 3$ kpc scales, and in \S 5 we present our
conclusions.

% DESCRIPTION OF OBSERVATIONS
\section{A New Survey for CO in Dwarf Galaxies}

\subsection{Sample Selection}

The original goal of this survey was to identify dwarf galaxies with
$^{12}$CO $J=1\rightarrow0$ emission suitable for high resolution
follow-up with the BIMA interferometer. Thus, the sample was
constructed to maximize the chances of detecting molecular emission.
We considered nearby ($V_{LSR} \lesssim 1000$ \kmpers), northern
($\delta > -5^{\circ}$), compact (optical diameter of $d_{25}< 5'$)
galaxies from the NGC, UGC, UGCA, IC, and DDO catalogs.  We included
only galaxies that show signs of ongoing star formation as indicated
by IRAS-detected 60 or 100 $\mu$m emission. We observed only dwarf
galaxies, which we defined as galaxies with an \hi\ linewidth
$W_{20}\lesssim 200$ km s$^{-1}$. Applying the Tully-Fisher relation
to an edge-on galaxy, this definition of dwarf corresponds to $M_B
\gtrsim -18$ \citep[$L_B \lesssim 3 \times 10^9 L_{\odot,B}$;
][]{SA00}. Finally, we removed galaxies that appeared tidally
disrupted or were obviously interacting with other galaxies. To
increase the observing efficiency, we added $18$ galaxies with no IRAS
emission to the target list in otherwise sparsely populated LST
ranges. These criteria produced a set of $152$ galaxies. Of these,
$121$ had no prior published \co\ detections. These $121$ galaxies
constitute our sample of targets.

\begin{figure*}
\epsscale{1.0}
\plotone{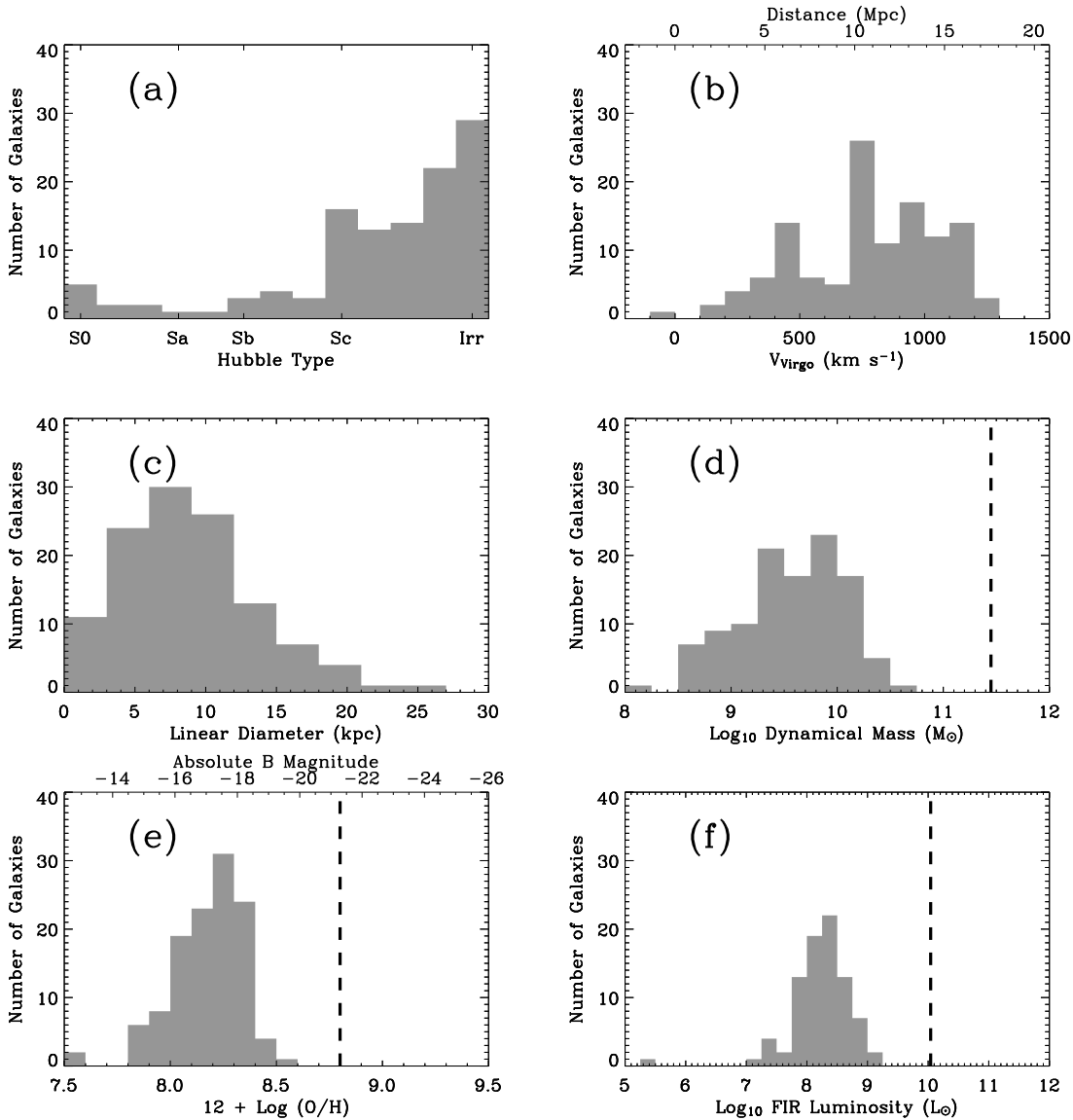} 
\figcaption{\label{SampleFig} A summary of some of the important
properties of the galaxies in our survey: (a) the Hubble Type; (b) the
recessional velocity, corrected for Virgocentric flow (equivalently
distance, shown for $H_0 = 72$ \kmpers\ Mpc$^{-1}$); (c) the linear
diameter in kpc, using the distances from (b) and $d_{25}$; (d) the
log of the dynamical mass, $M_{dyn} = v_{rot}^2 R / G$ (in solar
masses); (e) the metallicity, derived from the absolute blue magnitude
(shown along the top axis) using the relation of \citet{RM95}; and (f)
the FIR luminosity (Equation 2). Dashed lines show approximate values
for the Milky Way.}
\end{figure*}

Distributions of several key properties of the target galaxies are
displayed in Figure \ref{SampleFig}. The sample consists mostly of
late-type spiral and irregular galaxies at distances between 2 and 20
Mpc with typical linear diameters $\lesssim 10$ kpc and typical
dynamical masses $M_{dyn} = v_{rot}^2 R / G \lesssim 10^{10}$
M$_{\odot}$ (where $v_{rot}$ is the inclination-corrected maximum
rotation velocity obtained from \hi\ observations and taken from the
HyperLeda catalog \footnote{The HyperLeda catalog is located on the
World Wide Web at \\ \noindent {\tt
http://www-obs.univ-lyon1.fr/hypercat/intro.html}} and $R$ is the
optical radius at the 25 mag arcsec$^{-2}$ isophote assuming the
Virgocentric-flow corrected Hubble flow distances).

\subsection{ARO 12m Observations}

We observed our sample with the Arizona Radio Observatory (ARO) 12m
telescope at Kitt Peak. This telescope has a $55\arcsec$ half-power
beam-width at 115.27 GHz. We pointed toward the optical center of our
galaxies, obtained from the NASA/IPAC Extragalactic Database
(NED). The data were acquired over the course of three observing runs
(2001 February 1-7, 2002 January 21-25, and 2002 May 8-16) and several
nights of remote observing (2001 November 27-30, 2001 December 3, 2002
May 30-June 1). Whenever possible, we observed both polarizations with
the 1 MHz and 2 MHz filter banks in parallel mode, providing redundant
data on each polarization. For several runs this was not an option due
to hardware difficulties, and we used only one set of filter banks. We
observed each source for a minimum of one hour, divided into
six-minute integrations (scans). We allocated more time to sources
that showed some indication of emission after the first hour, sources
observed during poor weather, and those observed with only one
polarization. The observing mode was usually position switching, with
an offset of 2-3$'$ in azimuth.  In no case did we see evidence of
significant emission in the reference position close to the source
velocity, except for Galactic emission.  Every six hours, after
sunset, and sunrise, a planet or other strong continuum source was
observed to optimize the pointing and focus of the telescope. The
median system temperature taken over all runs was 345 K.

We reduced the spectrum for each six-minute scan in the following
manner. We removed noise spikes and bad channels by flagging all
channels with absolute values above the $5\sigma$ level (none of our
sources were this bright in a single scan). Several channels were
known to be bad {\it a priori} and we flagged these as well. We then
subtracted a linear baseline from the spectrum and binned it to our
final resolution of $10$ \kmpers. Finally, we averaged both
polarizations and all scans to produce the final spectrum for each
source. In very few ($8$) cases this procedure was not sufficient to
produce a useful final spectrum, and scans with very poor baselines
were either discarded or fit with a higher order polynomial.

\subsection{Detection Algorithm and Integrated \co\ Intensity}

We classified each galaxy as ``detected,'' ``marginally detected,'' or
``not detected'' based on the region of the spectrum containing the
most statistically significant emission. We selected this region by
applying the following simple algorithm to each of the final
spectra. Each channel (10 \kmpers\ velocity bin) within half of the
\hi\ linewidth, $W_{20}$, of the systemic velocity was used as the
center of a series of spectral windows ranging from 30 to 190 \kmpers\
in width (in 20 \kmpers\ increments). For each of these spectral
windows, we calculated the signal--to--noise ratio of the integrated
intensity as

\begin{equation}
SNR = \frac{1}{\sqrt{N} \mbox{ } RMS} \sum_{i=1}^N I_i,
\end{equation}

\noindent where $I_i$ is the brightness temperature (intensity) in the
$i$th channel within the window, $N$ is the number of spectral
channels in the window, and $RMS$ is the root mean square intensity of
the spectrum per channel measured in the (assumed) signal-free areas
outside $\pm 100$ \kmpers\ of the systemic velocity. For each
spectrum, we calculated $SNR$ for all of the spectral windows that met
the above criteria. With 9 spectral windows (widths of 3, 5, 7,
... 17, and 19 channels) centered on each channel within $W_{20}/2$
(typically $\sim 70$ \kmpers) of the systemic velocity, we examined
$\sim 135$ spectral windows per spectrum. Of these 135 values, we
selected the one with the maximum signal-to-noise ratio. We refer to
the SNR of this spectral window as the $SNR_{Max}$ of the spectrum.

We used a Monte Carlo technique to estimate the false positive rate of
this algorithm --- i.e., the probability of randomly generating a
$SNR_{Max}$ higher than a given value from spectra containing only
Gaussian noise. We generated a large number of artificial spectra
filled with normally distributed values and containing the same number
of velocity channels as our real spectra. We found that our algorithm
extracted $SNR_{Max}$ values of $4.3$ or higher from these artificial
spectra $< 0.5\%$ of the time, indicating $\sim 99.5\%$ confidence
that $SNR_{Max} \geq 4.3$ corresponds to real emission. Based on our
Monte Carlo results, we classified all spectra for which we found a
$SNR_{Max}$ of $4.3$ or higher as ``detections.''  Further, we found
that our algorithm extracted a $SNR_{Max}$ of $3.0$ or higher from the
artificial spectra 5\% of the time. We labeled spectra with a
$SNR_{Max}$ higher than $3.0$ but less than $4.3$ ``marginal
detections.''  These thresholds, $3.0$ and $4.3$, are dependent on the
number of independent channels in a spectrum and are therefore
specific to our data.  Given our sample of $121$ galaxies, we expect
$< 1$ false detection and $\sim 6$ false marginal detections. In
total, we find $28$ detections (none of which we expect are false) and
$16$ marginal detections ($\sim 6$ of which we expect are false).
Note that we omit the marginal detections from the analysis presented
in \S 3 and 4. However, we consider the majority of these galaxies
likely to be detected in \co\ by future studies.

We also estimated our false negative rate --- i.e., the probability
that our algorithm will fail to identify a spectrum known to contain
signal as a detection. We added signal with the median characteristics
of our typical detections ($\sigma_v = 50$ \kmpers, $I_{CO} = 1.5$ K
\kmpers) to an average noise spectrum ($RMS = 0.008$ K in a 10
\kmpers\ channel). Our algorithm recovered this signal as a detection
$\sim 70\%$ of the time, as a marginal detection $\sim 25\%$ of the
time, and not at all $\lesssim 5\%$ of the time. Under the simplifying
assumption that detected galaxies represent a uniform population, our
recovery rate and our $28$ detections imply a total population of $28
\times \frac{1}{0.7} \sim 40$ ``detectable'' galaxies in our sample.

We calculated the integrated intensity of each galaxy in the following
manner. For detections and marginal detections we considered the
spectral window that produced the maximum signal-to-noise value. When
this region was bordered by channels containing emission, we extended
the window to include all contiguous channels with positive
intensities. The integrated intensity of the galaxy in \co\ was taken
to be the sum over this spectral window, and the quoted error is the
statistical uncertainty over the same region. For non-detections, we
measured the statistical uncertainty in the integrated intensity over
a 120 \kmpers\ window (the \hi\ velocity width of a typical
nondetection) centered on the radio LSR velocity of the galaxy.

\begin{figure*}
\epsscale{0.9} 
\plotone{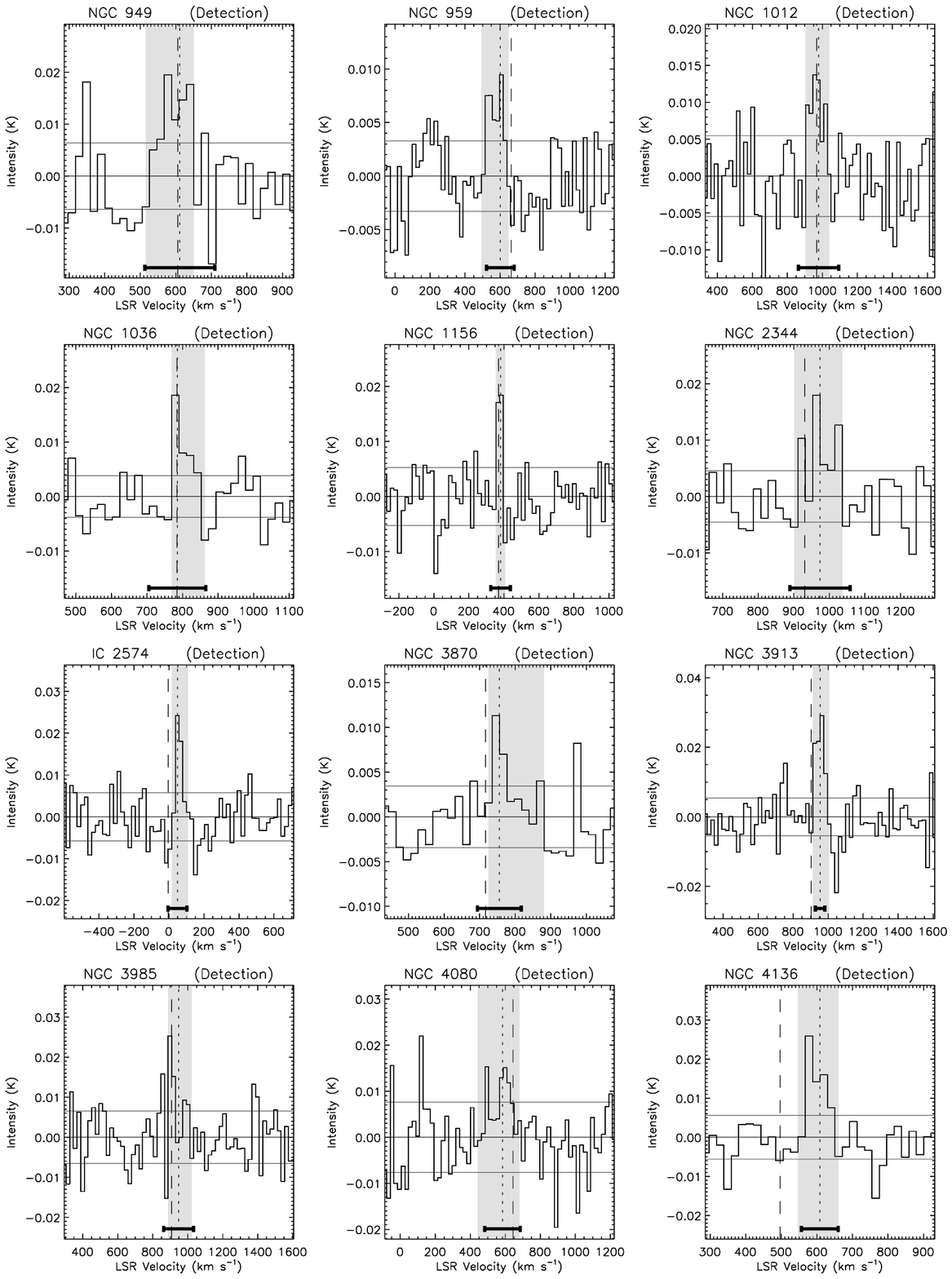} 
\figcaption{\label{detfig} Spectra of detected galaxies from the
survey. The horizontal lines indicate the RMS noise calculated from
the signal-free region of spectrum and the vertical lines show the
position of the optical (dashed) and radio (dotted) systemic
velocities. The shaded region is identified by our algorithm as
emission and the black bar shows the \hi\ linewidth, $W_{20}$. We
expect $\lesssim 1$ false positive among the 28 detections and $\sim
5$ false positives among the 16 marginal detections (see \S2).}
\end{figure*}

Table \ref{dettable} summarizes the survey results. Column (1) lists
the name of the galaxy; columns (2) and (3) give the center position
from NED; column (4) gives the LSR velocity of the galaxy (usually
derived from the \hi); column (5) lists the extinction-corrected
absolute $B$ magnitude of the galaxy, derived as described below;
column (6) contains the $\log_{10}$ of the FIR luminosity of the
galaxy (for galaxies that lack either 60 $\mu$m or 100 $\mu$m
emission, we do not report $L_{FIR}$); column (7) gives the 1.4 GHz
flux associated with the central \co\ pointing; column (8) lists the
inclination-corrected rotation velocity of the galaxy; column (9)
gives the integrated intensity corrected for the main beam efficiency
($I_{CO} = \int T_{mb} dv$) derived from the \co\ spectrum, along with
$1\sigma$ error bars (for detected galaxies), or a $3\sigma$ upper
limit (for nondetections); finally, column (10) indicates whether we
detected \co\ in that galaxy --- ``Y'' for yes, ``N'' for no, and
``M'' indicates a marginal detection. Figure \ref{detfig} shows the
spectra of detections and marginal detections, with the selected
spectral window shaded. The optical- and radio-derived systemic
velocities are shown as vertical lines and the horizontal lines show
the RMS noise derived from the signal-free regions of the
spectrum. Beneath the spectrum the \hi\ velocity width, $W_{20}$, is
indicated by a dark bar. For the convenience of the reader, we note
the following conversion factors: at 115 GHz with a $55''$ (half
power) beam, 1 \Kkmpers\ corresponds to a flux of 32.9 Jy \kmpers.
Assuming a Galactic \co-to-H$_2$ conversion factor \citep[$2 \times
10^{20}$ \xcounits, e.g. ][]{SM96}, this flux is equivalent to a
molecular surface density of 4.4 $\cos i$ M$_{\odot}$ pc$^{-2}$
(including helium), where $i$ is the inclination of the galaxy in
question. We shall consider variations in the conversion factor in
\S4.

\begin{figure*}
\epsscale{0.9} \figurenum{\ref{detfig}} \plotone{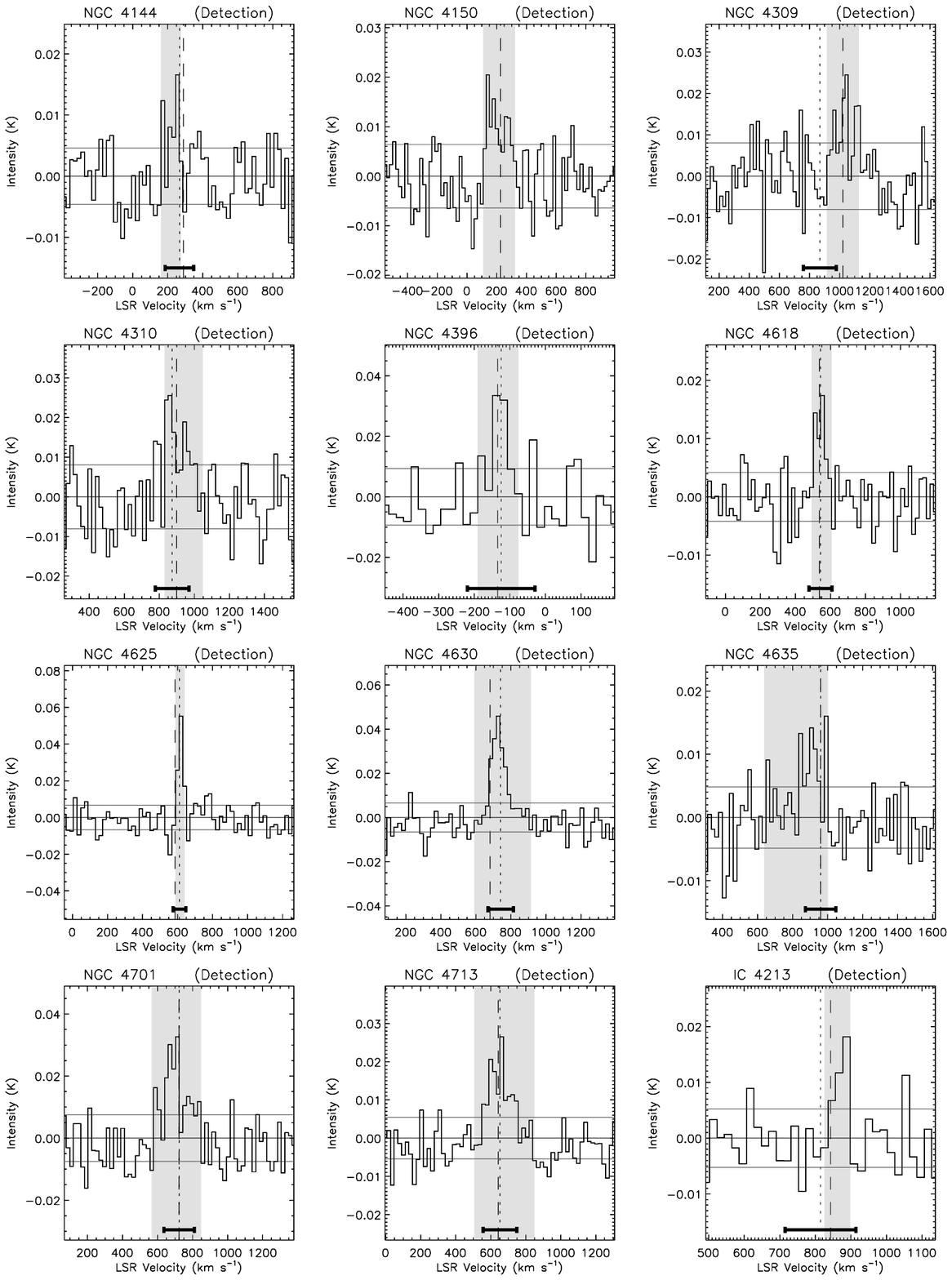}
\figcaption{Spectra of detected galaxies from the survey. The
horizontal lines indicate the RMS noise calculated from the
signal-free region of spectrum, and the vertical lines show the
position of the optical (dashed) and radio (dotted) systemic
velocities. The shaded region is identified by our algorithm as
emission and the black bar shows the \hi\ linewidth,
$W_{20}$. We expect $\lesssim 1$ false positive among the 28
detections and $\sim 5$ false positives among the 16 marginal
detections (see \S2). (continued)}
\end{figure*}

\begin{figure*}
\epsscale{0.9} \figurenum{\ref{detfig}} \plotone{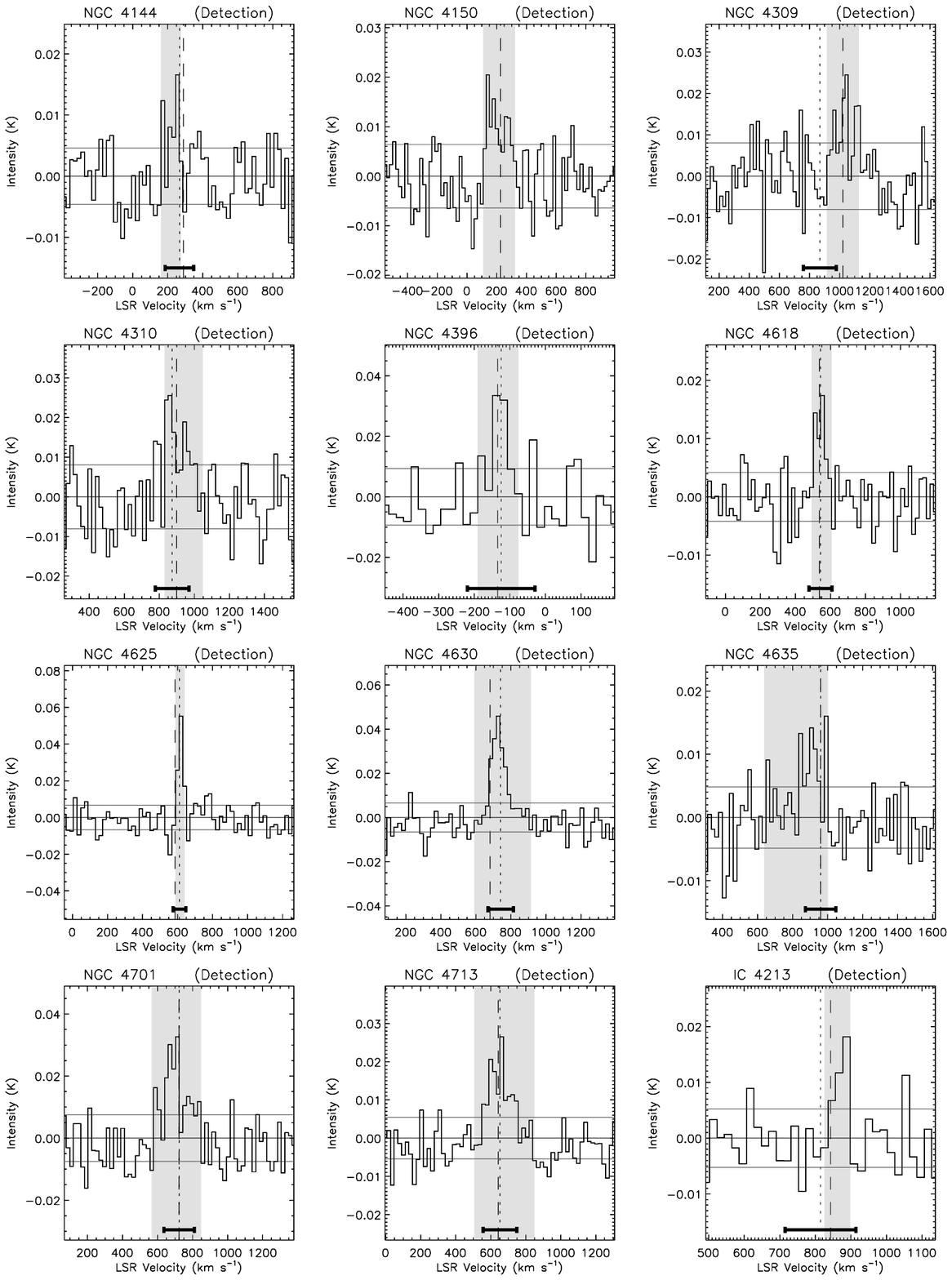}
\figcaption{Spectra of detected galaxies from the survey. The
horizontal lines indicate the RMS noise calculated from the
signal-free region of spectrum, and the vertical lines show the
position of the optical (dashed) and radio (dotted) systemic
velocities. The shaded region is identified by our algorithm as
emission and the black bar shows the \hi\ linewidth,
$W_{20}$. We expect $\lesssim 1$ false positive among the 28
detections and $\sim 5$ false positives among the 16 marginal
detections (see \S2). (continued)}
\end{figure*}

\begin{figure*}
\epsscale{0.9} \figurenum{\ref{detfig}} \plotone{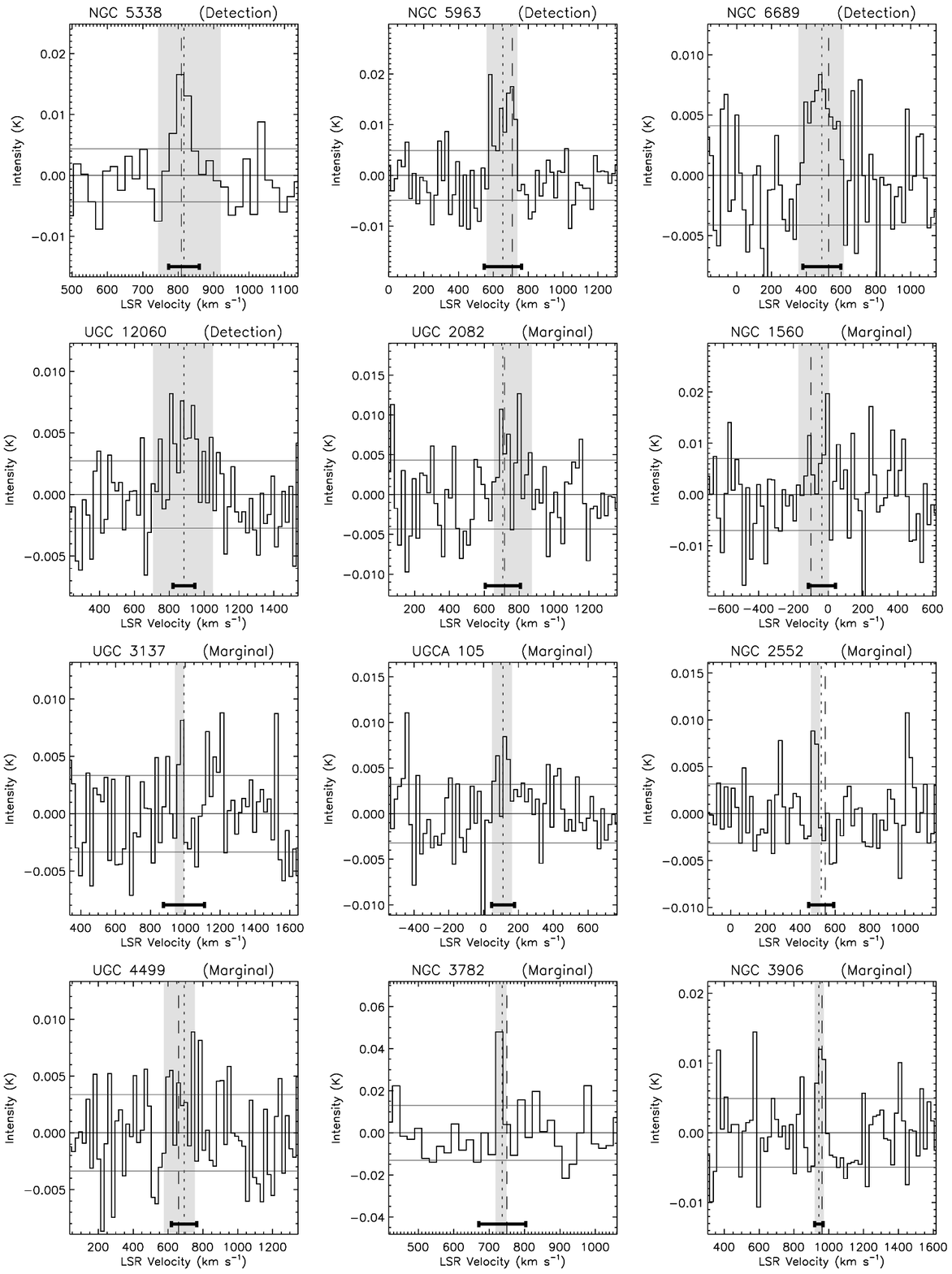}
\figcaption{Spectra of detected galaxies from the survey, followed by
those of marginally detected galaxies. The horizontal lines indicate
the RMS noise calculated from the signal-free region of spectrum and
the vertical lines show the position of the optical (dashed) and radio
(dotted) systemic velocities. The shaded region is identified by our
algorithm as emission and the black bar shows the \hi\ linewidth,
$W_{20}$. We expect $\lesssim 1$ false positive among the 28
detections and $\sim 5$ false positives among the 16 marginal
detections (see \S2). (continued)}
\end{figure*}

\begin{figure*}
\epsscale{0.9} \figurenum{\ref{detfig}} \plotone{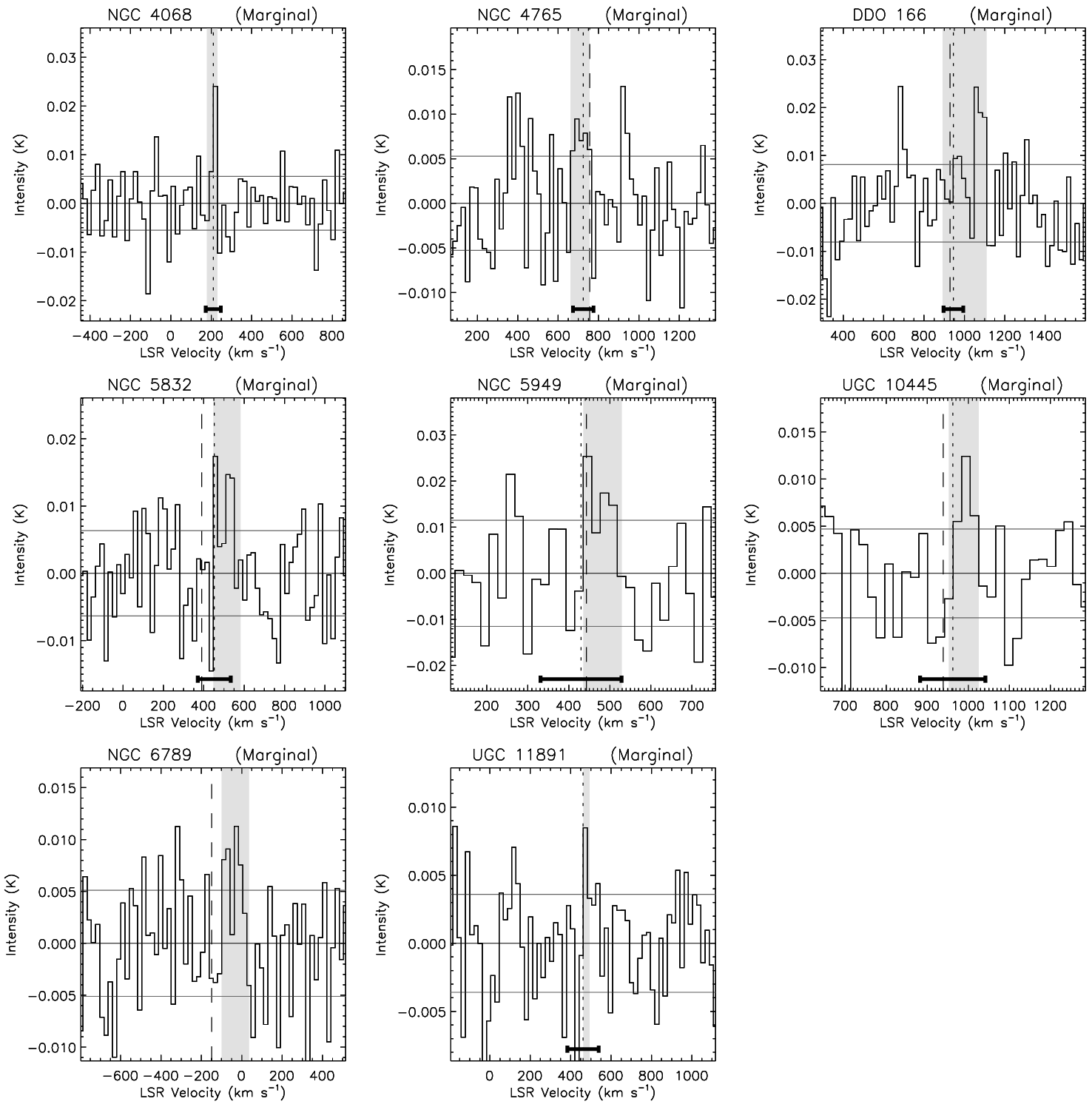}
\figcaption{Spectra of marginally detected galaxies from the
survey. The horizontal lines indicate the RMS noise calculated from
the signal-free region of spectrum and the vertical lines show the
position of the optical (dashed) and radio (dotted) systemic
velocities. The shaded region is identified by our algorithm as
emission and the black bar shows the \hi\ linewidth, $W_{20}$. We
expect $\lesssim 1$ false positive among the 28 detections and $\sim
5$ false positives among the 16 marginal detections (see
\S2). (concluded)}
\end{figure*}

\subsection{Supplementary Data}

To create a dataset that could be used to relate \co\ to other galaxy
properties, we supplemented our survey with data from the
literature. The \co\ data came from four sources: the observations and
compilation of \citet*[][]{TKS98} (using the ARO 12m telescope), the
FCRAO extragalactic \co\ survey \citep[][using the FCRAO 14m
telescope]{YO95}, the survey of late-type spirals by \citet[][]{BO03}
(using the IRAM 30m telescope), and the survey of \citet[][]{EH96}
(using the SEST 15m telescope and the Onsala 20m telescope). We refer
to this set of additional \co\ observations as the ``supplemental
sample.'' Note that the supplemental sample, in particular the FCRAO
survey but also the other surveys, contains a number of galaxies that
do not meet our definition of dwarf galaxy (i.e., for which $v_{rot} >
100$ \kmpers). These galaxies make up the sample of large galaxies
used for comparison in \S 3 and \S 4. We pared the literature sample
slightly, removing galaxies with Hubble types earlier than S0 and
galaxies with inferred molecular masses greater than their dynamical
masses. We consider either the molecular mass or the supplementary
galaxy properties of such galaxies unreliable.

We obtained systemic and rotational velocities, optical magnitudes,
21-cm fluxes, inclinations, diameters, and Hubble types from the
HyperLeda catalog. To these data, we added 1.4 GHz radio continuum
measurements (RC), far infrared (FIR) fluxes, and $K$-band magnitudes
from other sources. The RC measurements are measured from the NRAO VLA
Sky Survey (NVSS) images \citep[][]{CO98}, which have a resolution of
$45''$ (well matched to the $55''$ 12m beam) and a sensitivity of
$0.45$ mJy beam$^{-1}$. For each galaxy we calculated the RC flux
integrated over the entire disk of the galaxy, which is used as a
global property in \S 3. We also calculated the RC flux associated
with the \co\ observation from the NVSS convolved to the resolution of
that \co\ observation (when possible) for use in \S 4.

We computed the FIR flux \citep[][]{dV91},

\begin{equation}
\label{FIRDEF}
F_{FIR} = 1.26 \left[ 2.58 f_{\nu} (60) + f_{\nu} (100) \right] \times
10^{-14} \mbox{ W m}^{-2} \mbox{,}
\end{equation}

\noindent using IRAS 60 and 100 $\mu$m flux densities (in Janskys)
from the IRAS Faint Source Catalog \citep[][]{IRASFSC}. $F_{FIR}$
approximates the flux in a bandpass with uniform response 80 $\mu$m
wide centered at 82.5 $\mu$m. Since thermal emission from a galaxy
usually peaks between 50 and 100 $\mu$m, $F_{FIR}$ is a good indicator
of its total infrared flux \citep[usually about 50\%, see][]{BELL03}.

The $B$-band measurements used in this paper begin as apparent,
uncorrected $B$ magnitudes from the HyperLeda catalog. We correct for
internal extinction following \citet[][]{SA00}, and for Galactic
extinction using \citet[][]{DUST98}. We use the Virgocentric-flow
corrected Hubble flow distance with $H_0=72$ \kmpers Mpc$^{-1}$
\citep[][]{WF01}. One galaxy in our sample, NGC~4396, has a negative
distance using this method and we assigned it a distance of 20 Mpc
(under the assumption that it is a member of Virgo Cluster).

We took $K$-band measurements from the 2MASS Extended Source Catalog
(XSC) \citep[][]{JAR00}. We used the magnitudes obtained by
extrapolating the surface brightness profile to the extrapolated
K-band radius of the galaxy ($K_{ext}$ in the 2MASS XSC) because they
better recover the extended flux of low surface brightness objects. In
several cases the extrapolated K-band radius of the galaxy was smaller
than $d_{25}/2$, which prompted us to discard the $K$-band magnitude
for that galaxy as unreliable.

% DISCUSSION I: CO AND GALAXY PROPERTIES IN DWARFS
\section{CO Emission and Galaxy Properties}

In this section we examine the relationship between \co\ emission and
other galaxy properties. We find that the strongest correlations are
between the \co\ and the FIR, RC, $K$-band, and $B$-band
luminosities. The correlation between \co, FIR, and RC presumably
results from the well established association between molecular gas
and star formation, but the strong correlation between \co\ and
stellar light is somewhat surprising. It suggests that dwarf galaxies
are small versions of large galaxies and that differences in the \co\
content among galaxies are primarily a result of scaling --- a
galaxy's mass seems to be the main factor in setting its molecular gas
content. We argue below that this strong correlation between molecular
gas content and stellar luminosity arises from the crucial role played
by the gravity of the stars in setting the midplane gas pressure and
thus the local density of atomic gas --- which governs the rate of
H$_2$ formation.

\subsection{Molecular Gas in Dwarf Galaxies with Detected CO Emission}

We estimate the strength of the relationship between \co\ content and
other galaxy properties using the rank correlation coefficient.  The
magnitude of the rank correlation coefficient gives a robust, unbiased
measure of how strongly two variables are correlated but no indication
of the specifics of that relationship beyond the sense (obtained from
the sign). It is calculated in a manner analogous to the linear
correlation coefficient, but it uses the rank of a data point within
the sorted data in place of the actual value of the data point
\citep[for more details see][]{NR92}. The rank correlation coefficient
can therefore be used to distinguish whether an association exists
between two galaxy properties, but contains no information on the
physics of that association. To estimate the uncertainties in the
coefficients we use bootstrapping methods. We create an uncorrelated
dataset out of our data by randomly pairing $X$ and $Y$ data values,
and then we calculate the $1\sigma$ deviation of the rank correlation
coefficient around zero from many repetitions of the experiment.

Throughout this discussion we refer to the molecular mass, $M_{Mol}$,
rather than the observable $L_{CO}$, as the molecular mass is the
physical quantity of interest.  We calculate $M_{Mol}$ by applying a
Galactic CO-to-H$_2$ conversion factor \citep[$X_{CO} = 2 \times
10^{20}$ \xcounits;][]{SM96} to the \co\ luminosity. To accurately
account for the mass associated with molecular gas, we apply an
additional factor of 1.36 to include the effect of helium (see the
discussion in \S 4 for more details). Note that even if $X_{CO}$
varies from galaxy to galaxy (see \S 4.3 for more discussion of this
issue), our results will still hold for the relationship between
$L_{CO}$ and other galaxy properties.

In some cases drawn from the literature, there are observations of
several fields in one galaxy. In these cases, we sum all of the
emission from the galaxy to derive $M_{Mol}$. We removed galaxies with
angular sizes greater than $5\arcmin$ but without observations of
multiple fields from the analysis because we considered the \co\ in
the central $1 \arcmin$ an unreliable predictor of the total
$L_{CO}$. Although $M_{Mol}$ is strictly a lower limit (because our
observations cover only the central few kpc of each galaxy), it is
likely that in dwarf galaxies the central $\sim3$~kpc (diameter)
encompassed by our survey beam contains most of the CO emission. The
detections taken from the literature are, on average, farther away and
therefore do an even better job of including most of the emission.

\subsubsection{Molecular Gas Content and Global Properties}

\begin{figure*}
\plotone{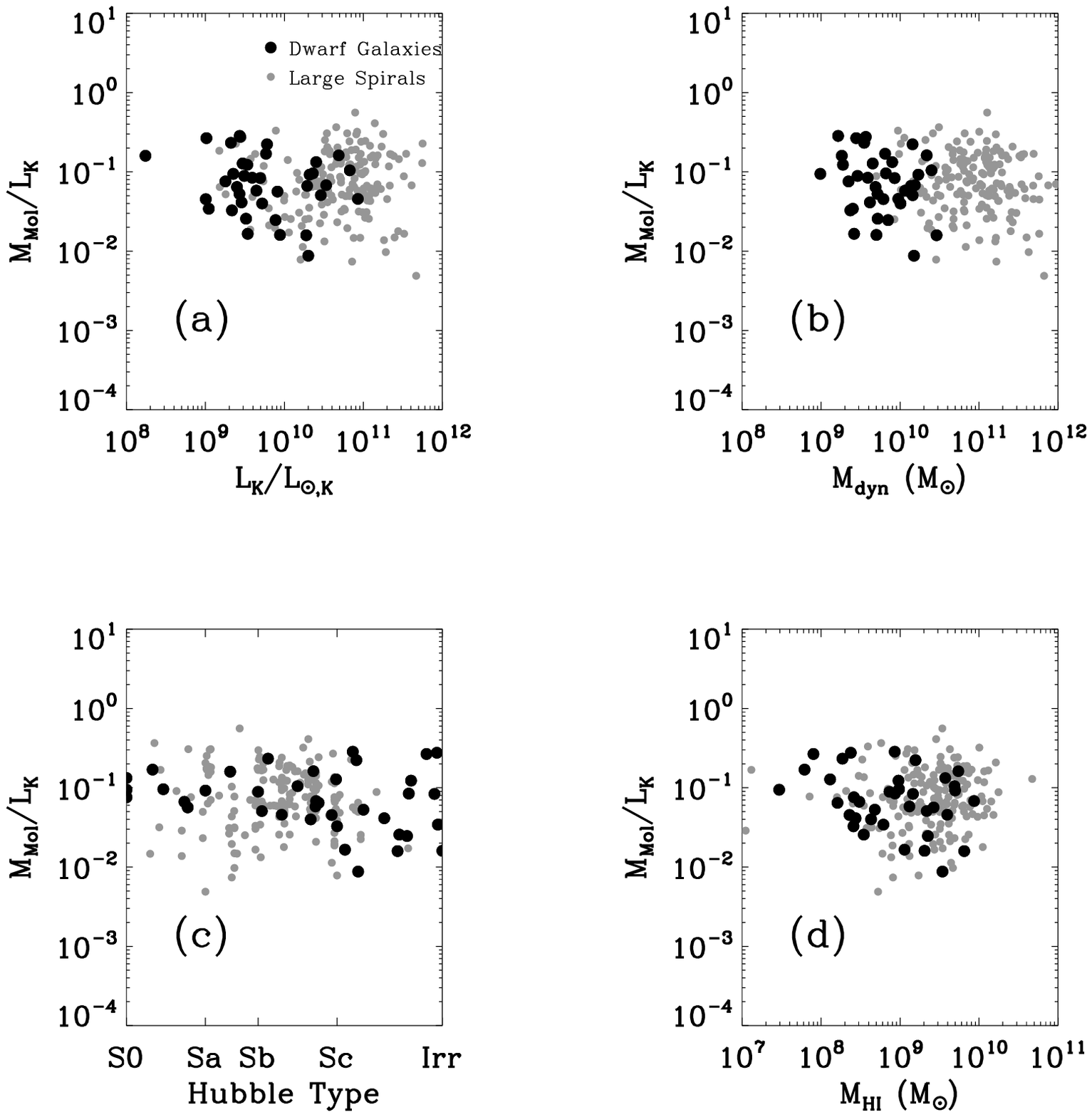}
\epsscale{1.0}
\figcaption{\label{COtoLKvsSIZE} Molecular gas mass per unit $K$-band
luminosity is remarkably constant as a function of galaxy mass and
morphology. Each panel shows $M_{Mol}/L_K$ as a function of a
different galaxy property. Panel (a) shows $K$-band luminosity, a
tracer of stellar mass; panel (b) show dynamical mass; panel (c) shows
galaxy morphology; and panel (d) shows atomic gas mass. In each case
dwarfs (black dots) and large galaxies (gray dots) show the same
median $M_{Mol}/L_K \approx 0.075 M_{\odot}/L_{\odot,K}$ and little or
no variation with the independent variable.  Dwarfs are defined as
galaxies with \hi\ inclination-corrected rotational velocities,
$v_{rot}$, less than 100 \kmpers.}
\end{figure*}

What properties of a galaxy are most important in setting its \co\
content? Are the same properties relevant in both dwarfs and large
spirals?  How much of the difference in the molecular gas content
between dwarfs and large spirals is a result of simple scaling with
galaxy size or mass?  In Table \ref{CorrTab} we present the answers to
some of these questions by showing rank correlation coefficients
between \co\ content --- both normalized and total --- and other
galaxy properties. The values shown make use of {\it all} galaxies in
our sample --- large spirals and dwarfs. By including both types of
galaxies we significantly extend the dynamic range over which we look
for variation (e.g., see Figure \ref{COtoLKvsSIZE}). Typical $1\sigma$
uncertainties in the rank correlation coefficients are $0.07$ and only
values with greater than $3\sigma$ significance are shown in the
table; other entries are left blank.

We also indicate in Table \ref{CorrTab} whether two quantities are
significantly correlated within the sample of dwarfs alone. Entries in
boldface indicate that the two quantities are correlated at $3\sigma$
significance or greater in the subset of dwarf galaxies. Typical
uncertainties in the rank correlation coefficients among the dwarfs
are $0.16$, so that boldfaced entries have rank correlation
coefficients $\gtrsim 0.5$ within the dwarf subset alone (and none
exceed $0.7$). In only one case, the correlation between $M_{HI}$ and
$M_{Mol}/M_{HI}$, do two quantities correlate significantly within the
dwarfs but not in the larger sample.

\begin{figure*}
\plotone{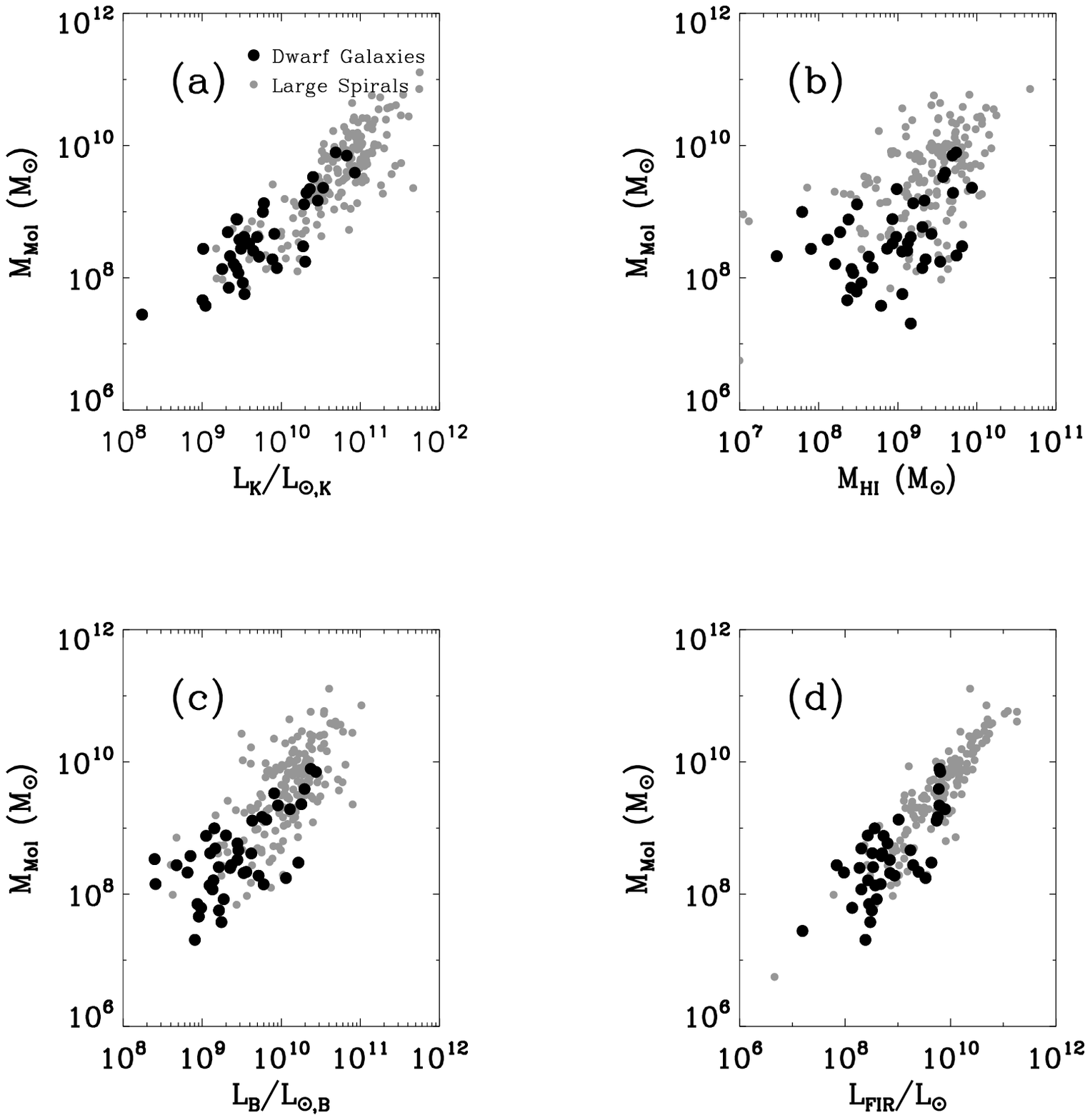}
\epsscale{1.0}
\figcaption{\label{COvsSIZE} Molecular gas mass as a function of
several galaxy properties: (a) $K$-band luminosity, (b) atomic gas
mass, (c) $B$-band luminosity, and (d) FIR luminosity.  Dwarfs are
defined as galaxies with inclination-corrected rotational velocities,
$v_{rot}$, less than 100 \kmpers.}
\end{figure*}

The second column in Table \ref{CorrTab} shows that $M_{Mol}$ is
correlated with a number of other galaxy properties at a very high
significance --- including Hubble type, dynamical mass, $K$-band
luminosity, $B$-band luminosity, FIR luminosity, RC luminosity, linear
diameter, and atomic hydrogen content. Figure \ref{COvsSIZE}
illustrates several of these correlations graphically, with dwarf
galaxies plotted as black circles and large galaxies as gray
circles. All of the correlations are in the same sense, namely that
more massive galaxies with earlier Hubble types and redder colors have
more molecular gas. In order to investigate the question of whether
dwarf galaxies are merely smaller version of large spirals, we must
look past scaling effects and examine how the {\it normalized} CO
content of a galaxy changes with its other properties.

\subsubsection{Normalized Molecular Gas Correlations} 

Columns (3) through (5) of Table \ref{CorrTab} show that many of the
strong correlations between CO content and other galaxy properties are
removed when the CO content is normalized by the stellar or dynamical
mass of the galaxy. In particular the quantity $M_{Mol}/L_K$ (column
3) shows very little systematic variation with other galaxy properties
(also see Figure \ref{COtoLKvsSIZE}). The median $M_{Mol}$/$L_K$ is
$\approx 0.075 \pm 0.005 M_{\odot}/L_{\odot,K}$ for large spirals and
$\approx 0.065 \pm 0.008 M_{\odot}/L_{\odot,K}$ for dwarf galaxies,
identical within the uncertainties --- Figure \ref{COtoLKvsSIZE} shows
this constancy. We find that similar results are achieved when
normalizing $M_{Mol}$ by the dynamical mass, $M_{dyn}$ (see column 4
of Table \ref{CorrTab}) --- $M_{Mol}/M_{dyn} \approx 0.040 \pm 0.003$
for large spirals and $0.037 \pm 0.007$ for dwarfs, again virtually
identical. The difference in $M_{Mol}$/$L_B$, which is $\approx 0.16
\pm 0.01 M_{\odot}/L_{\odot,B}$ for large spirals and $\approx 0.13
\pm 0.02 M_{\odot}/L_{\odot,B}$ for dwarf galaxies, is also not
significant. The weak, but significant, correlations between
$M_{Mol}$/$L_B$ and $L_K$, $B - K$, $M_{HI}/L_K$, and $M_{dyn}/L_K$
may reflect the fact that large galaxies tend to have redder stellar
populations than small galaxies. Also note that, although we apply an
extinction correction to the $B$-band luminosity that varies with
galaxy size \citep[][]{SA00}, extinction effects may still be
important.

These same results hold for a sample consisting of only dwarf
galaxies. $M_{Mol}$ in dwarfs correlates strongly with $L_K$, $L_B$,
and linear diameter. As in the larger sample, the $3\sigma$
correlations found within the dwarf galaxies disappear if the \co\
content is normalized by galaxy (stellar) mass. Thus we find little
evidence for systematic variations in the quantities $M_{Mol}/L_K$,
$M_{Mol}/M_{dyn}$, and $M_{Mol}/L_B$ with galaxy mass. This suggests
that dwarf galaxies are indeed ``small versions'' of large galaxies
and that differences in the \co\ content among galaxies are primarily
a result of scaling --- a galaxy's mass seems to be the main factor in
setting its molecular gas content and (stellar) mass alone can explain
most of the correlations seen in column (2) of Table \ref{CorrTab}.

\begin{figure}
\epsscale{1.2}
\plotone{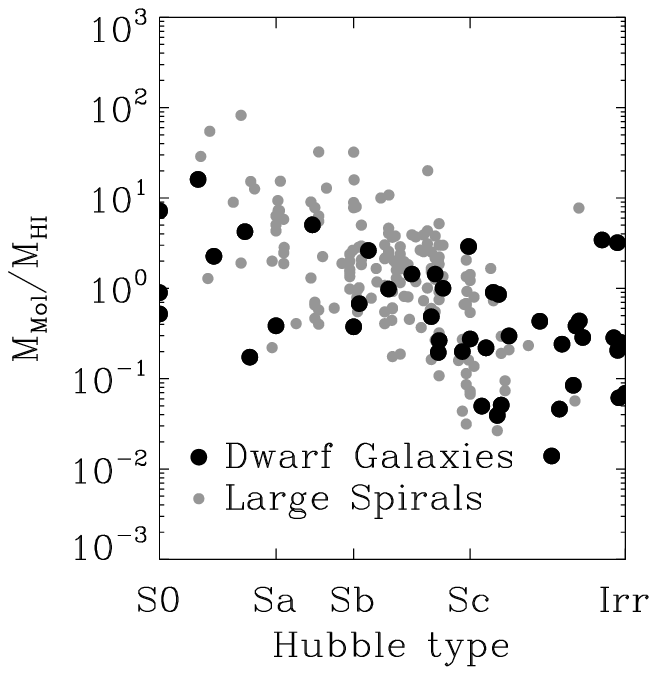}
\figcaption{\label{MOLFRACvsSIZE} The ratio of molecular to atomic gas
as a function of Hubble type. Later type galaxies, including most of
the dwarfs in our sample, have less molecular gas per unit atomic gas
than galaxies with earlier Hubble types. Dwarfs are defined as
galaxies with \hi\ inclination-corrected rotational velocities,
$v_{rot}$, less than 100 \kmpers.  The two irregular galaxies with
very large molecular fractions in Figure \ref{MOLFRACvsSIZE} are
NGC~4630 and NGC~4080. It seems quite possible that we have
overestimated the CO flux of NGC~4080 (see Figure \ref{detfig}), but
NGC~4630 clearly has much more CO emission than one would expect based
on its other properties.}
\end{figure}

The ratio of molecular to atomic gas, on the other hand, {\it is} a
strong function of other galaxy properties. Table \ref{CorrTab} and
Figure \protect\ref{MOLFRACvsSIZE} show the well-established trend of
decreasing $M_{Mol}/M_{HI}$ with later Hubble type and decreasing
galaxy mass \citep[see][and references therein]{YO91}. The dwarf
galaxies in our sample tend to be low mass systems with late Hubble
types and therefore have lower $M_{Mol}/M_{HI}$ than large spirals ---
$0.3 \pm 0.05$ compared to $1.5 \pm 0.1$. Although dwarf galaxies tend
to have roughly the same amount of molecular gas per unit stellar
mass, their ISMs are dominated by large reservoirs of atomic gas and
the molecular gas makes up only a small fraction of the total gas
mass.

\begin{figure*}
\plotone{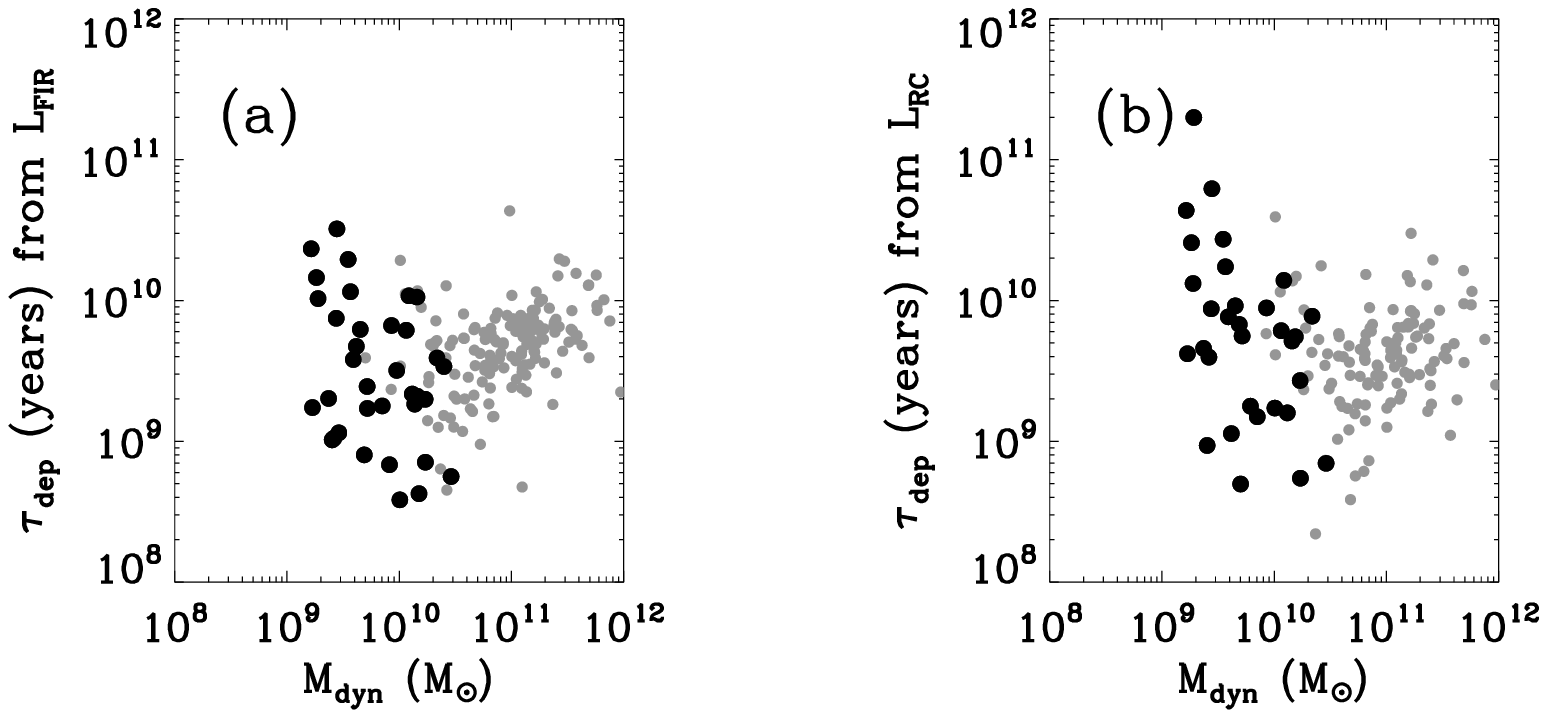}
\epsscale{1.0}
\figcaption{\label{SFEvsSIZE} Molecular gas depletion time decreases
with galaxy mass among dwarf galaxies (black dots) and then increases
with galaxy mass among large galaxies (gray dots). The two panels show
the same effect using different star formation tracers. Panel (a)
shows $\tau_{Dep}$ calculated using the high mass star formation rate
inferred from $L_{FIR}$ and panel (b) shows $\tau_{Dep}$ calculated
from the high mass star formation rate derived from the radio
continuum luminosity. Galaxies with masses $M_{dyn} \sim 10^{10}
M_{\odot}$ and appear able to convert their molecular gas into stars
most efficiently.  Dwarfs are defined as galaxies with \hi\
inclination-corrected rotational velocities, $v_{rot}$, less than 100
\kmpers.}
\end{figure*}

CO and FIR emission trace the amount of molecular gas and ongoing star
formation, respectively. Therefore the quantity $\tau_{Dep} =
M_{Mol}/L_{FIR}$ is a proxy for the depletion time, the time it will
take for a galaxy to consume its reservoir of molecular gas at its
present rate of star formation. Columns (7) and (8) of Table
\ref{CorrTab} and Figure \ref{SFEvsSIZE} show that $\tau_{Dep}$ varies
systematically with galaxy mass. Figure \ref{SFEvsSIZE} shows that
there is an anticorrelation between galaxy mass and the depletion time
within our sample of dwarfs. Figure \ref{SFEvsSIZE} also shows that
there is a positive correlation between mass and depletion time for
large galaxies. This latter result has been noted previously in a
careful study of face-on undisturbed galaxies by \citet[][]{YO99}. She
found that galaxies with diameters of $\sim 15$ kpc had the longest
depletion times in her sample (which did not include dwarfs), and that
the depletion times in larger galaxies increased with increasing
diameter.

Here we find that below masses of $\sim 10^{10} M_{\odot}$ and
diameters of $\sim 10$ kpc the depletion time tends to rise with
decreasing galaxy mass/size, so that galaxies with masses of $\sim
10^{10} M_{\odot}$ seem to be maximally efficient at turning molecular
gas into stars. Later in this paper, we find that molecular gas
depletion time decreases with increasing surface density of molecular
gas. That is, the higher the surface density of molecular gas the more
quickly it will be consumed by star formation. We suggest that this
effect drives the behavior in the dwarfs shown Figure
\ref{SFEvsSIZE}. Galaxies of $\sim 10^{10} M_{\odot}$ have higher
average surface densities of molecular gas than do smaller galaxies
and therefore they display slightly lower depletion times (i.e. higher
star formation efficiencies).

Based on our examination of the normalized molecular gas content we
find that the amount of molecular gas per unit stellar mass or
dynamical mass is roughly constant across more than two orders of
magnitude in galaxy mass. Over this same range, the molecular gas
depletion time does vary, but not particularly strongly. On the other
hand, smaller galaxies tend to be {\it much} richer in atomic gas than
large galaxies and therefore have much smaller fractions of their
total gas mass in molecular form than large galaxies. These results
suggest that the dwarf galaxies in our sample do look very much like
scaled down versions of large galaxies, with the major difference
being the presence of a reservoir of atomic gas in smaller systems ---
gas that is apparently irrelevant to star formation. Table
\ref{SlopeTab} shows the power law exponents derived from fits of
$M_{Mol}$ to several galaxy mass and star formation tracers. That the
slopes are all very close to unity reinforces the conclusion that
variations in $M_{Mol}$ may be just the result of mass scaling.

\subsubsection{The Molecular ISM of Dwarf Galaxies}

The strongest correlations in our sample (both considering the dwarfs
alone and the combined set of dwarfs and large spirals) are between
the molecular gas content, $M_{Mol}$, the $K$-band luminosity, $L_K$,
the $B$-band luminosity, $L_B$, and the FIR luminosity, $L_{FIR}$, of
galaxies. The correlation between $M_{Mol}$ and $L_{FIR}$ is well
known in large spiral galaxies \citep[e.g.,][]{YO91}: $L_{FIR}$ is
dust-reprocessed light from young stars, which have recently formed
out of the molecular gas. The relationship between $M_{Mol}$ and the
$K$-band luminosity (dominated by K-giants and other old stars --- a
proxy for the total stellar mass of a galaxy), is more puzzling. A
power law fit of $M_{Mol}$ to $L_K$ predicts the \co\ content of a
dwarf with less scatter ($\sim 0.4$ dex) than a similar fit to $L_B$
($\sim 0.5$ dex) or $M_{dyn}$ ($\sim 0.7$ dex). Further, the slope of
the best fit power law is nearly unity $1.3 \pm 0.2$ and normalization
by $L_K$ removes almost all correlation between $M_{Mol}$ and other
galaxy properties --- unlike normalization by $L_B$, which leaves weak
but significant correlation with $L_K$ (see Tables \ref{CorrTab} and
\ref{SlopeTab}).

Why should the correlation between molecular gas content and the mass
of old disk stars be as strong as that between molecular gas content
and the light of the young stars that form directly out of that gas?
Further, why should the correlation between molecular gas content and
the old disk stars be so much stronger than that between the molecular
gas mass and the mass of the atomic gas from which it forms? We
suggest that this correlation arises because the hydrostatic pressure
in the galactic disk regulates the rate at which H$_2$ forms out of
\hi, and that the stellar surface density drives the midplane
hydrostatic pressure.

Recently, \citet[][]{WO02} and \citet[][]{BR04} suggested that the
hydrostatic gas pressure plays a dominant role in setting the ratio of
atomic to molecular gas in disk galaxies. They show that the midplane
hydrostatic gas pressure, $P_h$, in a stellar-dominated galactic disk
can be written:

\begin{equation}
P_h\propto\sqrt{\Sigma_*}\,\Sigma_g\,\frac{\sigma_v}{\sqrt{h_*}},
\end{equation}

\noindent where $\Sigma_g$ is the surface density of gas in the disk,
$\sigma_v$ is the gas velocity dispersion, $\Sigma_*$ is the surface
density of the stars, and $h_*$ is the stellar scale height. There is
good evidence that $\sigma_v$ and $h_*$ are nearly constant within and
among disk galaxies and that radial variations in the surface density
of atomic gas, $\Sigma_{HI}$, are small compared to radial variations
in $\Sigma_*$ \citep[see references in][]{BR04,SV84,KVD02}. Therefore,
in the atomic-dominated regions of large disk galaxies, the portions
of these galaxies most similar to the dwarf galaxies in this paper,
the dominant variable setting the midplane gas pressure is
$\Sigma_*$. Indeed, in a sample of spiral galaxies, \citet[][]{BR04}
found that the transition from a molecule-dominated ISM to an
atomic-dominated ISM comes at a nearly constant stellar surface
density, $120$ M$_{\odot}$ pc$^{-2}$. If $P_h$ is set largely by
$\Sigma_*$ and the molecular gas abundance is determined by $P_h$,
this would explain the good correlation between $M_{Mol}$ and $L_K$
within our sample.

Why should $P_h$ control the conversion of \hi\ into H$_2$? The gas
pressure is given by $P_h = \rho_g \sigma_v^2$, so in a medium with
constant velocity dispersion, $\sigma_v$, $P_h$ is directly
proportional to the local gas density, $\rho_g$. The local gas density
sets the rate of H$_2$ formation, with formation $\propto \rho_g^2$
\citep[][]{HWS71}. In practice, the equilibrium abundance of H$_2$ may
depend more weakly on the local gas density, because the strength of
the dissociating radiation field is likely to be higher in regions of
high gas density.

Indeed, several observations suggest that lower $K$-band luminosities
of dwarf galaxies should correspond to lower average gas
densities. The average \hi\ surface density in the centers of dwarf
galaxies correlates well with their central stellar surface brightness
\citep[][]{SW02}, and the central stellar surface brightness increases
with increasing luminosity \citep[see e.g.][]{BL03}. Therefore we
expect higher gas surface densities, $\Sigma_g$, in higher luminosity
systems. Furthermore, shallower stellar potentials yield larger gas
scale heights, $h_g$, which translate into a further decrease in the
mean gas density $\rho_g \sim \Sigma_g/h_g$. Despite their large
reservoirs of atomic gas, smaller galaxies with less developed disks
and shallower stellar potential wells lack the gas densities necessary
to efficiently convert \hi\ into H$_2$.

Thus, we suggest that the link between the $K$-band luminosity and the
molecular gas mass $M_{Mol}$ is the hydrostatic gas pressure, or
equivalently the local gas density. Because higher $K$-band
luminosities correspond to larger stellar surface densities, higher
values of $L_K$ lead to larger midplane pressures and correspondingly
higher gas densities, making the \hi$\rightarrow$H$_2$ conversion more
efficient.  This effect translates into a greater abundance of
molecular gas in more luminous $K$-band galaxies, resulting in more
star formation (see \S4 for a discussion of the relationship between
\co\ and star formation) and the young, massive stars thus produced
will fuel the FIR luminosity of the galaxy.

While consistent with our data, this picture is not the only possible
interpretation. An alternative conclusion might be that in dwarf
galaxies (or actively star forming galaxies) the $K$-band light is a
good tracer of the star formation rate. For instance,
\citet[][]{RV94}, in a near-IR study of M~33 (a galaxy that is
intermediate between our dwarfs and the large spirals) found that OB
stars may affect the near-IR emission of a galaxy by as much as
$50\%$. Two observations lead us to prefer the stellar potential well
explanation offered above. First, the $K$-band light appears to be a
better predictor of \co\ content than the $B$-band light, which should
be even more sensitive to the population of young stars than the
$K$-band light. Second, the ratio of $M_{Mol}/L_K$ remains constant
across a range of morphologies (see Figure \ref{COtoLKvsSIZE}c) and
stellar populations --- in earlier-type galaxies the $K$-band light is
certainly not dominated by young stars and $M_{Mol}/L_K$ seems to be
constant among these systems and late-type, lower mass dwarfs. This
suggests that $L_K$ traces an older stellar population throughout our
sample.

\subsection{The CO Nondetections}

\begin{figure*}
\plotone{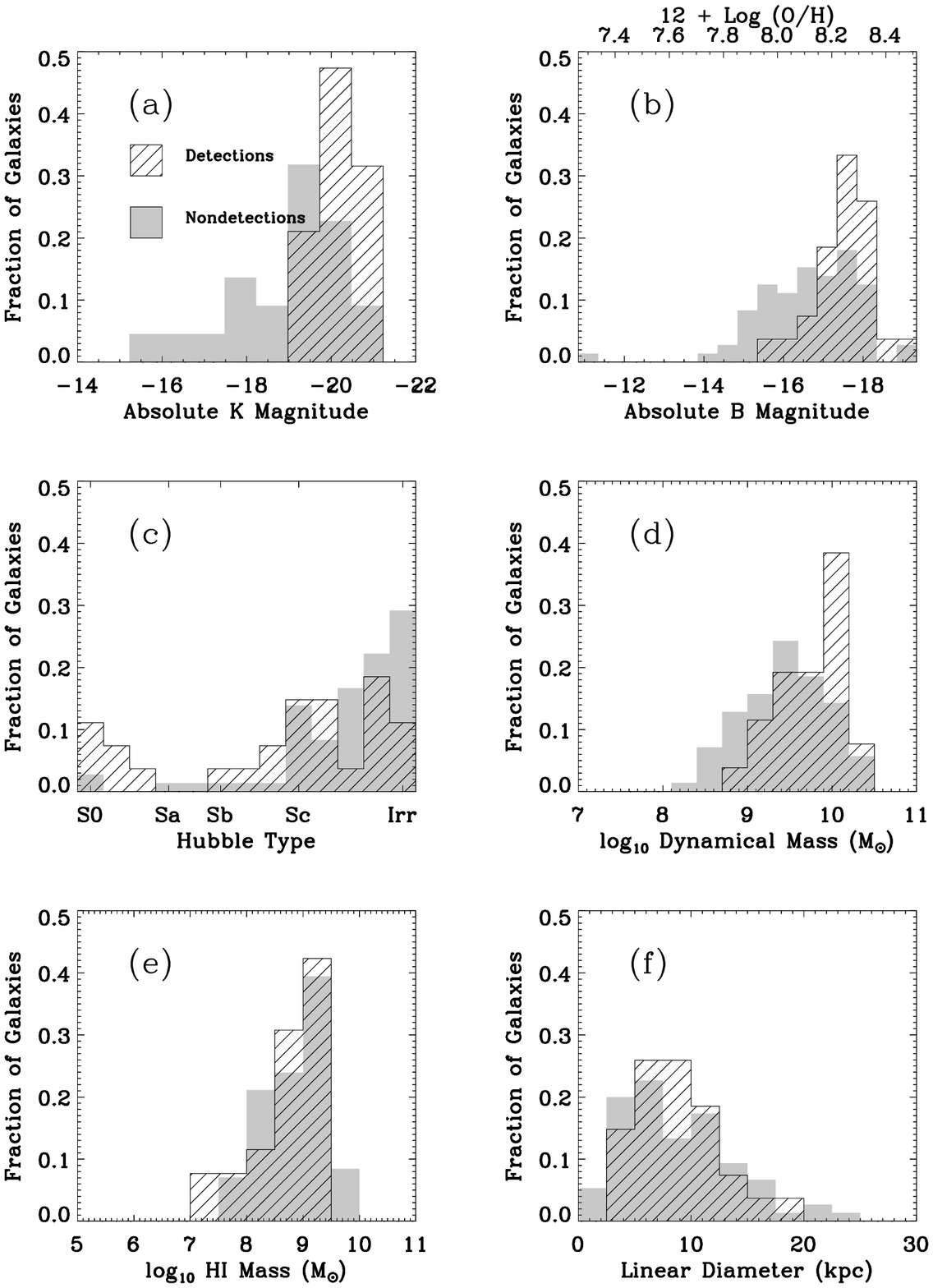} 
\epsscale{1.0}
\figcaption{\label{comphist1} Histograms showing the difference
between detections (hashed) and nondetections (filled) in
distributions of various parameters tracing galaxy size.}
\end{figure*}

\begin{figure*}
\plotone{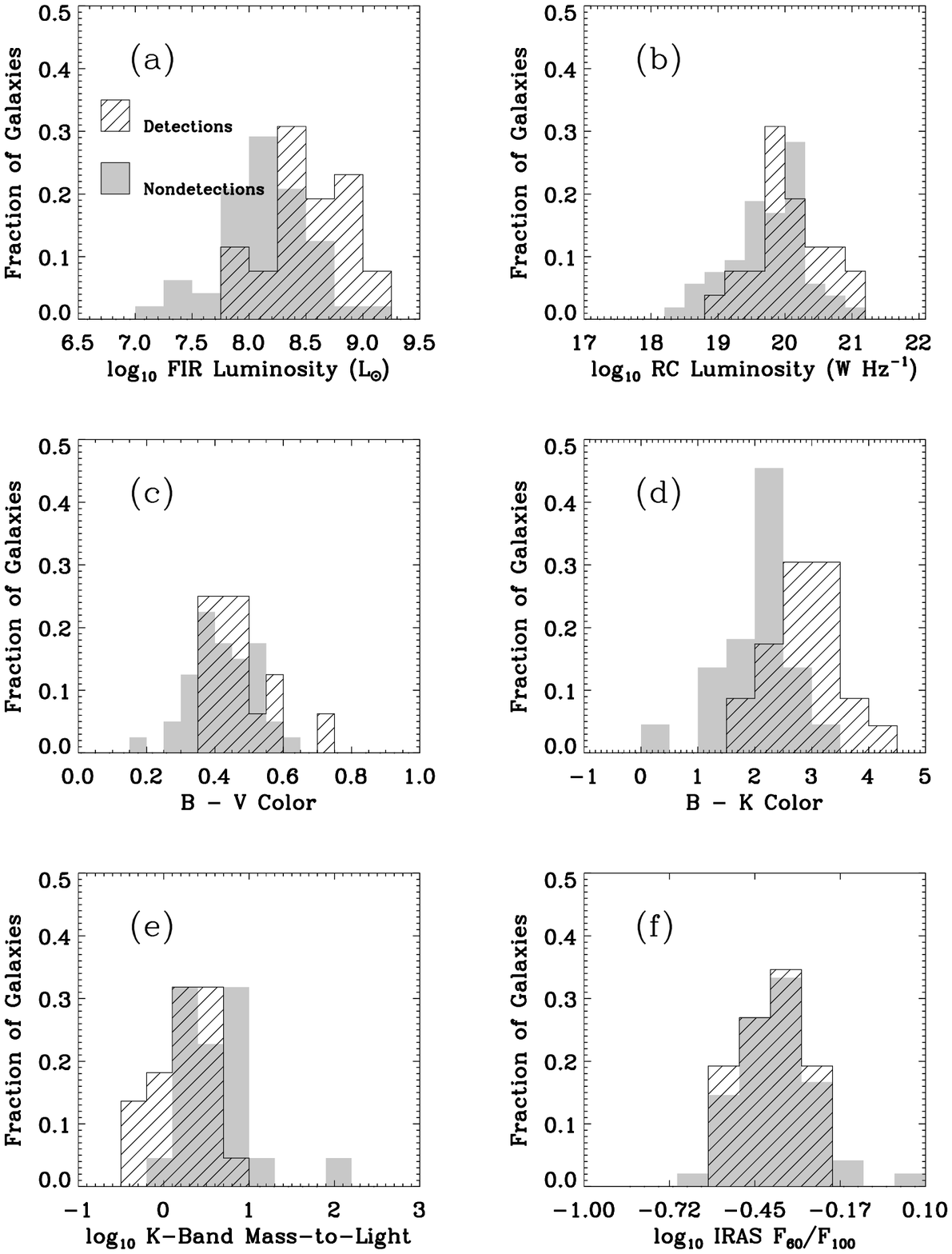}
\epsscale{1.0}
\figcaption{\label{comphist2} Histograms showing the difference
between detections (hashed) and nondetections (filled) in
distributions of star formation tracers, optical and infrared colors,
and mass-to-light ratios.}
\end{figure*}

How does this picture mesh with the properties of our \co\
nondections?  Here we restrict our investigation to our survey, a well
defined sample including detections and nondetections. We looked for
statistical differences between selected properties of galaxies
classified as detections and those of nondetections. Table
\ref{MedTab} gives median values for nondetections and detections for
each property and Table \ref{DiffTab} summarizes the significance of
these differences. Figures \ref{comphist1} and \ref{comphist2} show
the populations of detections and nondetections for a number of
properties of interest.

We calculated the significance of the differences in Table
\ref{MedTab} in the following manner. For each property, we used two
statistics to test the hypothesis that the samples of detected and
nondetected galaxies were drawn from the same parent
population. First, we used a two-sided Kolmogorov-Smirnov (KS) test,
which considers the maximum difference between the cumulative
probability distributions of the two populations. Second, we applied
the Student's $t$ test, which considers the difference between the
means of the two distributions. To estimate the significance of the
results we used a Monte Carlo approach, where we applied the tests to
randomly selected subsamples of the observed galaxies. Significant
measures of the differences between detected and nondetected galaxies
are those that are unlikely to be obtained from randomly selected
subsamples. The numbers listed in Table \ref{DiffTab} are the fraction
of randomly generated pairs of populations that produce a greater
(more significant) difference than we observe between the detected and
nondetected galaxies under the KS and Student's $t$ test, so that very
small numbers indicate very significant differences between the
detections and nondetections for that property.

Table \ref{DiffTab} shows that, not surprisingly, the distributions of
most of the galaxy properties seen to be strongly correlated to \co\
in Table \ref{CorrTab} also differ quite significantly between
detections and nondetections --- the most notable exception being the
linear diameter, which fails to distinguish between detections and
nondetections. The populations of absolute K magnitudes and absolute B
magnitudes differ between the detections and nondetections at greater
than $\sim 99\%$ significance in the sense that brighter galaxies are
more likely to be \co\ detections. Our two tracers of global star
formation rate, FIR luminosity and radio continuum luminosity, show
the same trend --- higher SFR increases the chances of detecting CO
(although this effect is of only marginal significance for
$L_{RC}$). As in Table \ref{CorrTab}, the total \hi\ mass associated
with the system seems to be a relatively poor indicator of \co\
content. Nondetections and detections have approximately the same
amount of atomic hydrogen, so that the nondetections are, in fact,
more gas rich (measured by $M_{HI}/L_B$) than the detections.

Thus, the picture outlined in the previous subsection appears to apply
to the \co\ nondetections in our survey. Galaxies with low stellar
masses (traced by the $K$-band and $B$-band luminosities) tend to be
nondetections even though they have comparable reservoirs of atomic
gas to the \co\ detections. We suggest that this is because the lower
surface density in the stellar disks of the nondetections leads to a
more diffuse, low-pressure distribution of atomic gas, one that is
consequently less suitable for converting atomic gas into molecular
gas.

\subsubsection{Portrait of a Detection and a Nondetection}

There is considerable latitude within the classification of a galaxy
as a ``dwarf,'' so here we paint a rough picture of the types of
galaxies that we detect in our survey and the types of galaxies that
remain beyond our detection limits.

Our detections tend to be dwarf spirals similar in luminosity and mass
to the LMC, though typically richer in \co\ and forming stars somewhat
less vigorously. Their median dynamical mass and B-magnitudes
($M_{dyn} = 8 \times 10^{9}$ M$_{\odot}$ and $M_B = -17.6$) are
comparable to those of the LMC \citep[which has $M_{dyn} \approx 8
\times 10^{9}$ M$_{\odot}$ and $M_B \approx -17.8$;][]{K98}. Most of
our detections an order of magnitude greater \co\ luminosity than the
LMC \citep[$\approx 2 \times 10^6$ K km s$^{-1}$ pc$^2$;][]{MIZ01}
within our $\sim3$~kpc beam. Although most of the \co\ emission in the
LMC is concentrated within a $\sim 3$ kpc area, this area is not at
the optical center of the galaxy. Therefore this luminosity is
actually an upper limit to what we would expect to detect if an LMC
analog were included in our sample, and about a fifth of detected
galaxies show \co\ luminosities similar to or less than this
value. Assuming a Galactic CO-to-H$_2$ conversion factor, the median
CO mass for a detection is $8 \times 10^7$ M$_{\odot}$, about 10\% of
the median \hi\ mass. This molecular fraction is about order of
magnitude lower than what is typical in large spirals. Most of our
galaxies have earlier Hubble types (median Sc) than the LMC (Sm). The
median FIR luminosity of a detection is $\sim 3 \times
10^{8}~L_{\odot}$, about half that of the LMC.

The nondetections are typically less massive and less luminous than
the detections and have later Hubble Types. Both their median
dynamical mass and their median optical luminosity ($M_{dyn} = 3
\times 10^9$ and $M_B = -16.8$) are similar to that of the SMC
($M_{dyn} = 2.4 \times 10^9$ M$_{\odot}$ \citep[][]{SS04} and $M_B =
-16.6$) and the Local Group galaxy IC~10 ($M_B = -16.4$). The median
FIR luminosity of a nondetection is $1.5 \times 10^{8}~L_{\odot}$,
which is several times higher than that of the SMC ($4.4 \times
10^{7}~L_{\odot}$) or IC~10 \citep[$3.6 \times
10^{7}~L_{\odot}$][]{MI94}. This should not be surprising, since we
required a galaxy to be detected by IRAS in order to be included in
our sample. The galaxies that we do not detect, then, are dwarf
irregulars with stellar masses similar to the SMC though with
significantly more vigorous star formation.

Thus, the molecular gas in very primitive systems, the distant cousins
to the SMC, remains tantalizingly out of reach. \citet[][]{MIZ01B}
found a \co\ luminosity of $8 \times 10^4$ K km$^{-1}$ pc$^2$ for the
SMC (though they did not survey the entire galaxy). If all of this gas
were concentrated within the 12m beam, it would yield an intensity
(averaged over the beam) of $\sim 1$ K km s$^{-1}$ at 1 Mpc, slightly
lower than the median intensity of one of our detections ($1.5$ K km
s$^{-1}$). At the distance of our nearest nondetection (2.2 Mpc), the
SMC drops to $\sim 0.25$ K km s$^{-1}$, which is undetectable by our
observations (the lowest intensity for which we detect \co\ in this
sample is $0.5$ K km s$^{-1}$). At the 11 Mpc median distance to a
member of our sample, this intensity drops to $\sim 0.01$ K km
s$^{-1}$, far below our detection limits. The SMC, however, is too
faint in the FIR to have made it into our IRAS-selected sample at such
a distance.

\subsubsection{FIR Color}

Does the size and temperature of the dust grains affect the \co\
emission from a galaxy? Dust helps to shield the gas from dissociating
UV radiation and is believed to provide the sites for H$_2$
formation. Variations in either the size distribution or temperature
of the dust in a galaxy might have an important effect on the
molecular gas content of that galaxy, especially since the range of
temperatures at which molecular gas can form on dust grains may be
quite small. Using the IRAS measurements at 12, 25, 60, and 100 $\mu$m
we looked for systematic variations in the FIR colors of our galaxies
that might be correlated with molecular gas content.

Galaxies detected in \co\ are also detected by IRAS at 12$\mu$m and
25$\mu$m at a much higher rate than \co\ nondetections (2 of the 74
nondetected galaxies have associated IRAS 12$\mu$m emission and only
11 are detected at 25$\mu$m; by comparison, 10 of the 28 detected
galaxies show 12$\mu$m emission and 11 show 25$\mu$m emission). It is
tempting to interpret this comparative lack of 12$\mu$m emission among
the nondetected galaxies as arising from a preferential depletion of
small grains analogous to that seen in the SMC
\citep[][]{SA90}. However, the upper limits associated with the
12$\mu$m emission in the \co\ nondetections are high enough that these
galaxies could have the same $F_{12}$/$F_{100}$ ratio as the \co\
detections and still not appear in the 12 $\mu$m catalog. Thus a
deficiency of small grains is not ruled out, but there is no strong
evidence for it. In addition, \co\ nondetections in which 25 $\mu$m
emission is seen have the same median $F_{25}$/$F_{100}$ ratio as the
\co\ detections ($\sim 0.05$).

The $F_{60}$/$F_{100}$ ratio, an indicator of the temperature of dust
in these galaxies, also shows no significant variation between
detections and nondetections ($F_{60}$/$F_{100} \approx 0.40$ for
nondetected galaxies and $0.41$ for detected galaxies). Further,
column (2) in Table \ref{CorrTab} shows that there is no significant
correlation between FIR color and total molecular gas content. Within
the combined sample of dwarfs and large spirals, $F_{60}$/$F_{100}$
{\it does} correlate with the normalized molecular gas content of a
galaxy. Galaxies with higher $F_{60}$/$F_{100}$ ratio tend to have
more molecular gas per unit mass/luminosity than less
``infrared-blue'' galaxies. Columns (7) and (8) also show that such
galaxies also have shorter depletion times. This is a result of the
fact that $F_{60}$/$F_{100}$ is an excellent predictor of $L_{FIR}$
for a given $L_K$ (or other mass indicator) --- $F_{60}$/$F_{100}$ and
$L_{FIR}/L_K$ are correlated at $\sim 10\sigma$, hardly surprising
given the definition of $L_{FIR}$ (see Equation \ref{FIRDEF}). Since
there is not even a suggestion of these trends within the dwarf
population alone, the usefulness of the far infrared color as a
predictor of CO emission {\it within dwarfs} appears to be
negligible. If a trend is present, the scatter among galaxies is too
large for it to be useful.

\subsubsection{Metallicity}

It has long been thought that the abundance of heavy elements in a
galaxy may be closely linked to its \co\ emission
\citep*[][]{MB88,TKS98}. Here we look for such an effect in our
sample. However, the number of well-determined metallicities for dwarf
galaxies remains small. Fortunately, within galaxies which do have
known metallicities there is a strong correlation between the $B$-band
luminosity of a dwarf galaxy and its metallicity \citep[][]{RM95}. In
fact, in dwarf irregulars the $B$-band luminosity is a better
predictor of metallicity than the central surface brightness
\citep[see][and references therein]{SK03}. The fit,

\begin{equation}
\label{RMZeq}
12 + \log \mbox{(O/H)} = (5.67 \pm 0.48) + (-0.147 \pm 0.029) M_B
\end{equation}

\noindent is derived from Local Group dwarf irregulars
\citep[][]{RM95} but also describes Sculptor Group dwarf irregulars
well \citep[][]{SK03} and the slope agrees within the uncertainties
with that found for dwarf galaxies using the best quality data in the
2dF Galaxy Redshift Survey \citep[][]{LA04}. For luminosities fainter
than $M_B = -15$ there is significant scatter, usually towards higher
than predicted abundances, but only a few of our galaxies --- all
nondetections --- are this faint.

\begin{figure}
\plotone{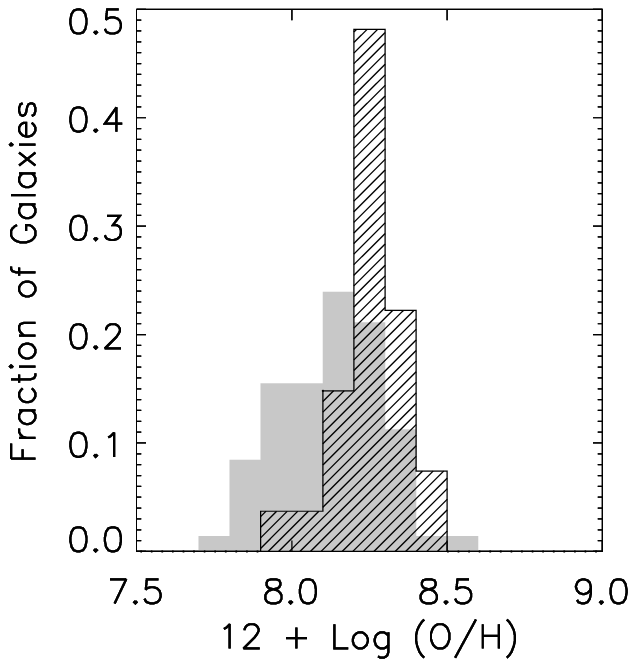} 
\epsscale{1.0}
\figcaption{\label{zfig} Metallicities (derived from absolute $B$-band
magntiudes) of dwarf galaxies in our survey. Nondetections (the solid
gray histogram) have systematically lower metallicities than
detections (the hashed histogram). Below $12 + \log \mbox{(O/H)} \sim
8.2$ the detection rate drops from $\approx 50\%$ to $\approx 20\%$.}
\end{figure}

Figure \ref{zfig} shows that there is significant difference in the
$B$-band derived metallicities of the detections and nondetections in
our survey. Above $12 + \log \mbox{(O/H)} = 8.2$, 21 of the 52
galaxies in our sample are detected (a further 3 are marginally
detected). Below that value only 6 of the 62 galaxies were detected
(and 9 are marginally detected). This drop in detection rate is
difficult to conclusively attribute to metallicity
effects. Metallicity is strongly correlated with other galaxy
properties --- indeed, we are using B magnitude as a proxy for
metallicity here --- and it is not clear whether a drop in detection
rates at low metallicities is caused by lower metallicities or by the
effects of other parameters that are covariant with
metallicity. Regardless, we do find agreement with \citet*[][]{TKS98}
who found that detections of galaxies with metallicities of $12 + \log
\mbox{(O/H)} \approx 8.0$ or less are quite rare. It remains to be
determined whether this is simply because such systems have less
molecular mass or because they form less \co\ per unit molecular mass.

 % DISCUSSION II: CO AND STAR FORMATION IN DWARFS
\section{Molecular Gas and Star Formation in Dwarf Galaxies}

We have already seen (in Table \ref{CorrTab} and the previous section)
that there is a strong association between molecular gas and star
formation --- traced by FIR and RC emission --- in our sample. Here we
investigate whether that association is the same in dwarf galaxies and
large spiral galaxies. Specifically, we ask: do dwarf galaxies differ
from large spiral galaxies in their mode of star formation, as
measured by the relationship between the surface density of molecular
gas, $\Sigma_{Mol}$, and the surface density of star formation,
$\Sigma_{SFR}$? There is reason to think that this might be the
case. Dwarfs differ from large spirals in a number of important
respects: luminosity, mass, metallicity, large scale dynamical effects
(such as spiral density waves and supernova-driven shocks), dust
properties, UV radiation field, and possibly magnetic field
strength. Each of these differences could conceivably change the
properties of molecular clouds in a manner that affects the rate at
which stars form out of molecular gas. Any such difference should
manifest itself as a different large scale relationship between
$\Sigma_{Mol}$ and $\Sigma_{SFR}$.

\subsection{The CO-to-RC Relationship in Dwarfs and Large Spirals}

\begin{figure*}
\epsscale{1.0}
\plotone{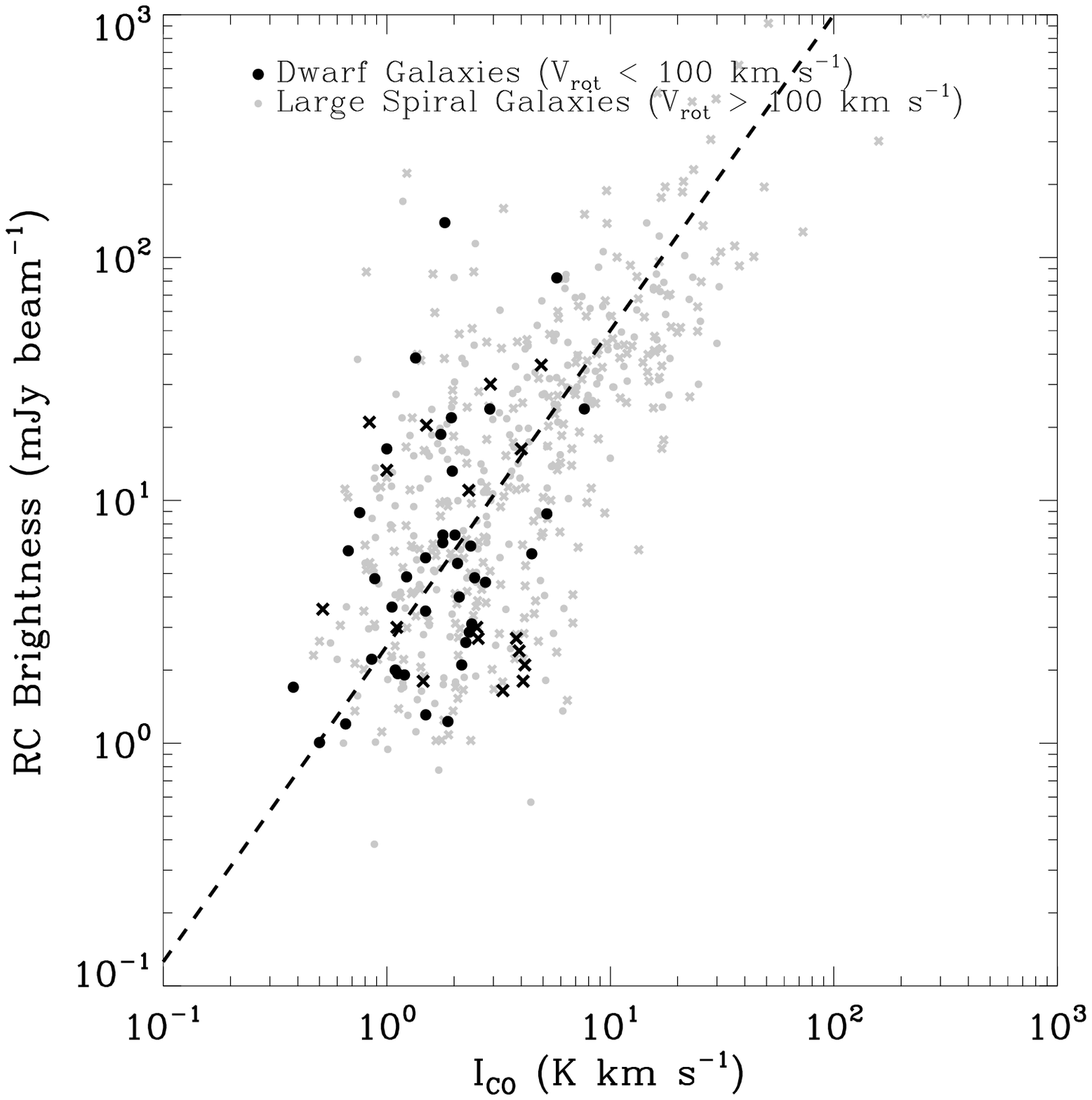} 
\figcaption{\label{RCvsICO} The 1.4 GHz radio continuum brightness
(flux density per $45''$ beam) plotted against integrated intensity
for dwarf galaxies (black points) and large spiral galaxies (gray
points). The dashed line shows the best fit to the two populations
combined. In contrast to Figure \ref{SFRvsH2}, all galaxies are
plotted here. Some galaxies were discarded from our analysis because
of their large angular sizes or early Hubble Types (see discussion in
\S4). These are plotted as crosses rather than circles and not used in
the fit.}
\end{figure*}

Radio continuum offers two major advantages over other star formation
tracers for the work in this section: 1) RC data are free of
extinction and 2) the entire sky is available in the NVSS at $45''$
resolution (comparable to the $55''$ 12m beam). This latter advantage
allows us to measure $\Sigma_{SFR}$ with the same resolution that we
use to measure $\Sigma_{Mol}$ ($\sim 3$ kpc for a typical dwarf
galaxy). Figure \ref{RCvsICO} shows the RC brightness (flux density,
in mJy, per $45''$ beam), $B_{\nu,1.4}$, and the integrated \co\
intensity, $I_{CO}$ (in K \kmpers), associated with \co\ pointings
from our survey and the supplemental sample (only galaxies with both
\co\ and RC detected at $\geq 2\sigma$ significance are included).

Figure \ref{RCvsICO} shows that $B_{\nu,1.4}$ and $I_{CO}$ are highly
correlated in both large spirals and dwarfs. This association has long
been known in large spirals \citep[][]{AD91,MU02} and we see here that
it extends seamlessly to dwarf galaxies. Indeed, we find that the
power laws that best describe the two datasets independently,

\begin{equation} 
B_{\nu,1.4}^{\mathrm{Large~Spirals}} ~ \left( \mathrm{mJy~beam}^{-1}
\right) = 10^{0.5 \pm 0.1} ~ I_{CO}^{1.4 \pm 0.2} ~ \left( \mbox{K
km s}^{-1} \right)
\end{equation}

 \noindent and

\begin{equation}
B_{\nu,1.4}^{\mathrm{Dwarfs}} \left(\mathrm{mJy~beam}^{-1}\right) =
10^{0.5 \pm 0.1} ~ I_{CO}^{1.5 \pm 0.2} ~ \left( \mbox{K km s}^{-1}
\right)
\end{equation}

\noindent agree within the uncertainties. {\em The simplest
interpretation for this agreement is that dwarfs and large spirals
show the same relationship between molecular gas and star formation
and that the \co\ and RC trace these physical quantities in the same
way in all galaxies}. In this context, dwarf galaxies appear to be
simple low mass mass versions of large spirals.

To arrive at the relationship between $\Sigma_{Mol}$ and
$\Sigma_{SFR}$, we use the following relations \citep[based on those
of][]{MU02} to convert RC surface brightness and $I_{CO}$ into the
physical quantities of interest,

\begin{eqnarray}
\label{RCtoSFR}
\Sigma_{SFR} \left(\frac{M_{\odot}}{\mbox{yr kpc}^2}\right) = 8.0
\times 10^{-4} \\ \nonumber \times ~ B_{\nu,1.4} \left(\mathrm{mJy~
beam}^{-1}\right) \,
\left({\frac{\theta}{\mbox{45\arcsec}}}\right)^{-2} \cos i
\end{eqnarray}

\begin{equation}
\label{COtoMol}
\Sigma_{Mol} \left( \frac{M_{\odot}}{\mathrm{pc^2}} \right) =
4.4~I_{CO} \left( \mbox{K km s}^{-1} \right) ~ \cos i \mbox{,}
\end{equation}

\noindent where $\theta$ is the linear size of the \co\ beam (in
\arcsec) or 45\arcsec, whichever is larger.  These equations assume
linear relationships between $B_{\nu,1.4}$ and $\Sigma_{SFR}$ for all
galaxies and a constant conversion factor of $X_{CO} = 2 \times
10^{20}$ \xcounits\ \citep[i.e., the Galactic value,][]{SM96}. We will
consider the effect of relaxing these assumptions in \S
\ref{xcosection}. Note that Equation \ref{COtoMol} includes a factor
of 1.36 to account for the presence of helium (hence we call it
$\Sigma_{Mol}$ rather than $\Sigma_{H_2}$). The purpose of this
convention is to account accurately for the mass of molecular clouds
and its conversion to stellar mass in \S \ref{SFefficiency}.

Our \co\ dataset is composed of observations with resolutions that
range from $24\arcsec$ to $55\arcsec$. For \co\ pointings observed
with a beam larger than $45''$, we measured the radio continuum
brightness from the NVSS over an area matching the \co\ beam (i.e., we
convolved the NVSS to the resolution of the \co\ observation). For
galaxies observed in \co\ with a beam smaller than $45''$, we used the
brightness in a single NVSS beam pointed at the center of the galaxy
(with a $45''$ area). The NVSS noise level is $0.45$ mJy beam$^{-1}$,
which translates into $\Sigma_{SFR} = 3.6 \times 10^{-4}$
M$_{\odot}$~yr$^{-1}$~kpc$^{-2}$ for a face-on galaxy.

\begin{figure*}
\epsscale{1.0} 
\plotone{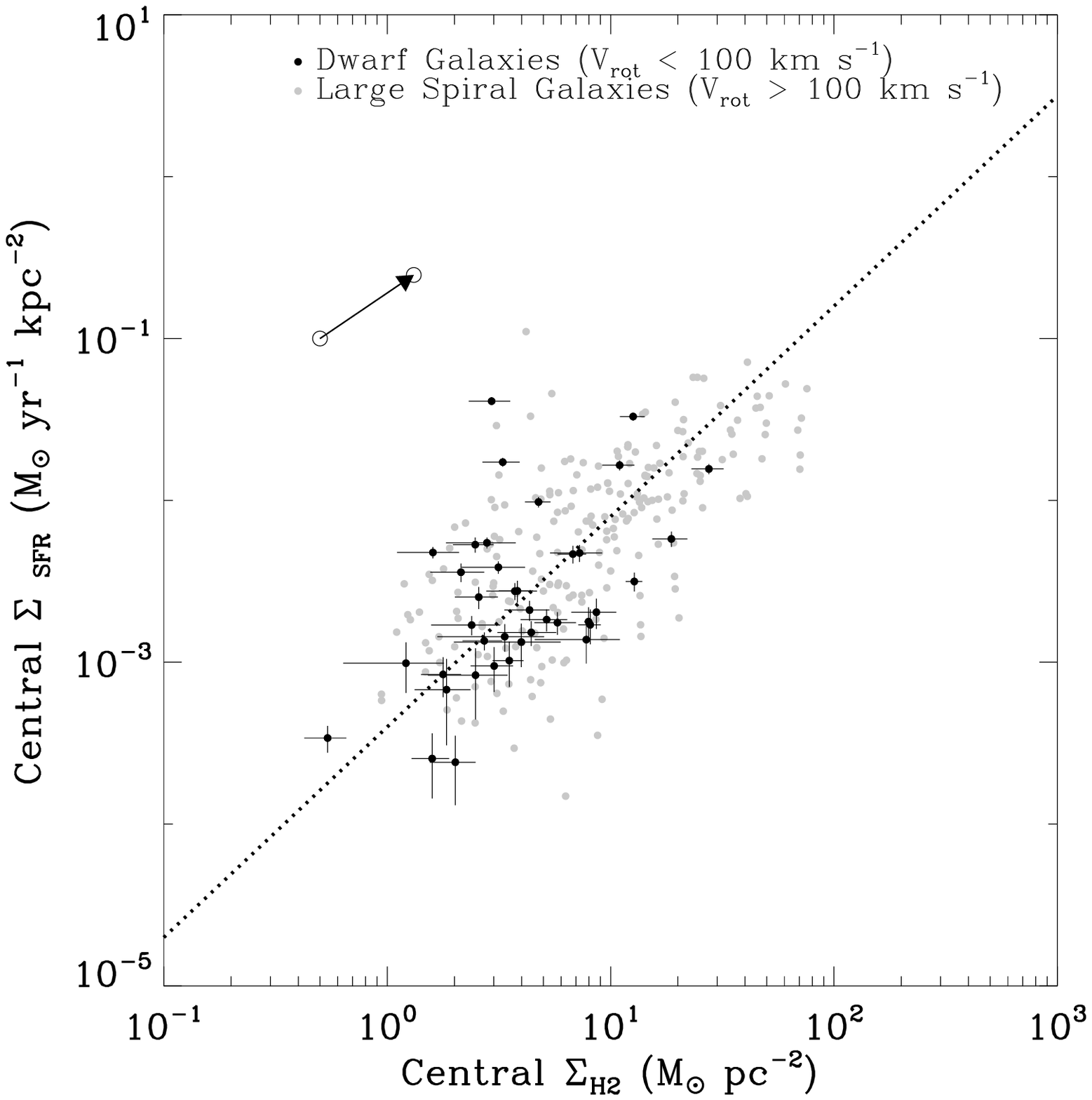}
\figcaption{\label{SFRvsH2} The star formation surface density of
spiral and irregular galaxies, derived from the 1.4 GHz radio
continuum, plotted against the central surface density of molecular
gas, derived from the integrated intensity of the CO line. Dwarf
galaxies (those with $V_{\mbox{rot}} \leq 100$ \kmpers) are displayed
as solid black circles, while large spiral galaxies are shown as gray
circles. The Schmidt Law that best fits all galaxies is shown as a
dotted line. The open circles connected by an arrow show the effect of
a typical correction using \citet[][]{WI95} and \citet[][]{BELL03} as
described in the text, but these corrections are not applied to the
data in this plot. The sample of galaxies in this figure has been
adjusted slightly from that in Figure \ref{RCvsICO}. To create a
homogeneous subsample free from spurious bulge contributions or
inclination effects, we discarded all galaxies with Hubble Types
earlier than Sb, inclinations $> 85^{\circ}$ and angular diameters $>
7'$.}
\end{figure*}

To create a homogeneous subsample free from spurious bulge
contributions or inclination effects, we discarded all galaxies with
Hubble Types earlier than Sb, inclinations $> 85^{\circ}$ and angular
diameters $> 7'$. We labeled galaxies with inclination-corrected
$v_{rot} < 100$ \kmpers\ dwarf galaxies and all others as large
spirals. The results below are fairly insensitive to this sample
selection. We find that changes in the criteria listed here have an
effect of no more than $0.1$ in the exponent and $0.2$ in the
coefficient. Using this subsample we determined that a ``Schmidt Law''
law of the form

\begin{equation}
\Sigma_{SFR} = a \left(\Sigma_{Mol}\right)^N
\end{equation}

\noindent is a good description of the data (Figure \ref{SFRvsH2}). To
quantify the underlying relationship between $\Sigma_{SFR}$ and
$\Sigma_{Mol}$, we followed the suggestion of \citet[][]{IS90} and
used the ordinary least squares (OLS) bisector method (the geometric
mean of the OLS fit of $X$ to $Y$ and that of $Y$ to $X$). We applied
bootstrapping techniques to estimate the robustness of the OLS
bisector fits. That is, we randomly drew a number of points from our
data equal to the number in the original sample, allowing points to
repeat and calculated the OLS bisector fit from each sample. We
repeated this process many times and estimated the uncertainty in each
parameter from the variation of fitted values. From these tests, we
conclude that the galaxies (large and dwarf combined) in our sample
are best described by the following power law:

\begin{equation}
\Sigma_{SFR} = 10^{-3.4 \pm 0.1} \Sigma_{Mol}^{1.3 \pm 0.1} \mbox{ ,}
\end{equation}

\noindent a relationship that agrees within the errors with that
derived by \citet[][note that there is significant overlap between our
sample and theirs]{MU02}:

\begin{equation}
\Sigma_{SFR} = 10^{-3.6 \pm 0.2} \Sigma_{Mol}^{1.3 \pm 0.1} \mbox{ ,}
\end{equation}

\noindent which we have adjusted to reconcile differences in the
assumed value of \xco\ and to account for the helium mass. If we
consider the dwarf and large spiral galaxies in our sample separately,
we find

\begin{equation}
\label{DwarfLaw}
\Sigma_{SFR}^{\mathrm{Dwarfs}} = 10^{-3.4 \pm 0.1} \Sigma_{Mol}^{1.3 \pm 0.2}
\end{equation}

\begin{equation}
\label{LargeLaw}
\Sigma_{SFR}^{\mathrm{Large~Spirals}} = 10^{-3.4 \pm 0.1}
\Sigma_{Mol}^{1.2 \pm 0.1}
\end{equation}

\noindent in good agreement with one another and with the results of
\citet[][]{MU02}. Thus, we find that dwarf galaxies display the same
relationship between \co\ and RC as large galaxies. As long as the
assumptions that lead to Equations \ref{COtoMol} and \ref{RCvsICO} are
valid, this result means that dwarf galaxies exhibit the same star
formation rate, $\Sigma_{SFR}$, and molecular gas depletion time,
$\tau_{Dep}$, for a given amount of molecular gas, $\Sigma_{Mol}$, as
large spiral galaxies. Thus, if the properties of molecular clouds do
vary with environment, they apparently do so in a way that does not
affect the large scale relationship between star formation and
molecular gas.

The slope derived here agrees within the uncertainties with that found
by \citet[][]{KE89}, who used the total gas surface density
($\Sigma_{\rm HI} + \Sigma_{H2}$) rather than the molecular gas
surface density. Dwarfs in our sample have ISMs dominated by atomic
gas, so the agreement between these two indices using different gas
surface densities bears investigating. However, we lack resolved
atomic hydrogen observations of most of the dwarfs in our sample and
therefore cannot further investigate the question of whether the
atomic/total or molecular gas surface density is most relevant to star
formation.

\subsection{Molecular Gas Depletion Times and Star Formation Efficiency}
\label{SFefficiency}

Equations \ref{DwarfLaw} and \ref{LargeLaw} can be adjusted to yield
the depletion time, $\tau_{Dep}$ for molecular gas as a function of
$\Sigma_{Mol}$:

\begin{equation}
\label{TauDep}
\tau_{Dep} = 2.5 \times 10^9 \Sigma_{Mol}^{-0.3} \mbox{ years}
\end{equation}

\noindent with $\Sigma_{Mol}$ measured in M$_{\odot}$ pc$^{-2}$. The
depletion time then is the time to consume all of the available
molecular gas at the present rate of star formation. Because
$\Sigma_{SFR}$ rises faster than $\Sigma_{Mol}$, the greater the
surface density of molecular gas, the faster the gas is depleted. The
median $\Sigma_{Mol}$ for a dwarf galaxy in our sample is $\sim 3$
M$_{\odot}$ pc$^{-2}$, which corresponds to a depletion time of
$\approx 1.8 \times 10^9$ years. For 10 M$_{\odot}$ pc$^{-2}$, a value
typical for large spirals, $\tau_{Dep}$ drops to $\approx 1.25 \times
10^9$ years. Strongly molecule-dominated galaxies, where
$\Sigma_{Mol}\sim300$ M$_{\odot}$ pc$^{-2}$, have
$\tau_{Dep}\sim4.5\times10^8$ years.

The star formation efficiency (SFE) is often used to characterize star
formation. The SFE can be defined as the fraction of molecular gas
turned into stars over $10^8$ years. In this case SFE is proportional
to the inverse of $\tau_{Dep}$ and equation \ref{TauDep} can be
rearranged to yield

\begin{equation}
SFE = 0.04 \Sigma_{Mol}^{0.3}
\end{equation}

\noindent with $\Sigma_{Mol}$ again in M$_{\odot}$ pc$^{-2}$. The SFE
of a typical dwarf ($\Sigma_{Mol} = 3$ M$_{\odot}$ pc$^{-2}$) is
$\approx 0.056$ and that of a large spiral ($\Sigma_{Mol} = 10$
M$_{\odot}$ pc$^{-2}$) is $\approx 0.08$.

Thus, although the two types of galaxy appear to obey the same
underlying relationships, this does not imply that they exhibit the
same SFE or $\tau_{Dep}$. Because dwarf galaxies tend to have lower
average $\Sigma_{Mol}$ than large spirals, they exhibit slightly lower
SFE and slightly larger $\tau_{Dep}$ than more massive systems. This
explains the trend of decreasing $\tau_{Dep}$ with galaxy mass seen in
Figure \ref{SFEvsSIZE}. The gradual upward trend among more massive
galaxies is more puzzling. If real, it could be a result of a slight
decrease in average molecular gas surface density with increasing
galaxy mass, but we see no strong evidence such a decrease among the
large galaxies in our sample. We also note that the depletion time for
both types of galaxies is $\sim 10^9$ years, much less than a Hubble
time. For star formation to be an ongoing process over a Hubble time
in all of these systems, molecular gas must be continuously
replenished.

\subsection{Effects of the CO-to-H$_2$ and RC-to-SFR Conversion
Factors}
\label{xcosection}

The results of the previous section are particularly surprising
because it has often been suggested that in low metallicity
environments like dwarf galaxies the CO-to-H$_2$ conversion factor,
\xco, may be much higher than the Galactic value
\citep[e.g.,][]{MB88}. A number of studies of dwarf galaxies using
virial methods have found values of \xco\ that are larger than the
Galactic value \citep[][and references
therein]{WI95,AST96,MIZ01,MIZ01B}. Far infrared studies also find
large variations in \xco\ with metallicity and radiation field
\citep[][]{IS97}. Recent applications of the virial method to nearby
galaxies at higher resolutions and signal-to-noise ratios, though,
find little evidence for changes in \xco\ with environment
\citep[][]{RO03,WA01,WA02,BOL03}. Though these new results for GMCs
are compelling, their applicability to clouds other than massive GMCs
is uncertain and the issue of the \co-to-H$_2$ calibration remains an
open one.

The data shown in Figures \ref{COtoMol} and \ref{RCvsICO} provides
some constraint on variations in \xco. If \xco\ increases with
decreasing metallicity, and the RC-to-SFR conversion does not change
with metallicity, then the (black) data points in Figure \ref{SFRvsH2}
will move from left to right. If we assume that the underlying
relationship between SFR and molecular gas is the same for large
spirals and dwarf galaxies, then any variations in \xco\ within the
dwarfs in our sample would have to be small in order to maintain the
agreement between dwarfs and large galaxies seen in Figure
\ref{RCvsICO}. Even if we allow the RC-to-SFR conversion vary with
metallicity, to hide a change in \xco\ of an order of magnitude over
the range of metallicities of our sample requires changing the
RC-to-SFR conversion by a similar factor, which appears very unlikely.

How much is the RC-to-SFR calibration likely to change over our
sample?  The conversion between RC and SFR relies on an empirical
calibration of the nonthermal synchrotron emission to the supernova
rate. Alternatively, the RC-to-SFR conversion can be calibrated via
the FIR-to-RC correlation, since the FIR luminosity of a galaxy can be
directly converted into a star formation rate once the amount of UV
light reprocessed by dust is known. Both of these methods will break
down when applied to sufficiently small objects. The fraction of
relativistic electrons that escape from a galaxy is not well known,
but cosmic rays may be more likely to escape from a small galaxy by
diffusion or convection \citep[e.g.,][ and references
therein]{CO92}. Therefore, a given amount of star formation could lead
to less RC emission in dwarfs than in large galaxies. Similarly, small
galaxies, with their shorter path lengths to escape and lower dust
abundances, are likely to reprocess less of the UV photons emitted by
young stars into FIR emission. Since the FIR and RC are strongly
correlated down to luminosities of $L_{FIR} \sim 10^8 L_{\odot}$ or
lower, if FIR traces SFR nonlinearly then RC must also trace SFR
nonlinearly \citep[][]{BELL03}. Although a thorough calibration effort
of the RC-to-SFR relation for small systems has yet to be undertaken,
this reasoning suggests that applying the RC-to-SFR calibration
derived in the Milky Way or other large galaxies to smaller systems
would yield SFRs that are too low.

To investigate the effects of possible changes in the CO-to-H$_2$ and
RC-to-SFR calibrations in our derived star formation laws we apply to
our sample some of the corrections suggested in the literature.  We
modified Equation \ref{RCtoSFR} to include a luminosity correction, as
suggested by \citet[][]{BELL03},

\begin{eqnarray}
\Sigma_{SFR} \left(\frac{M_{\odot}}{\mbox{yr kpc}^2}\right) = 8.0
\times 10^{-4} \\ \nonumber \times F_{\nu,1.4} ~ \left(\mbox{mJy}\right) \,
\left({\frac{\theta}{\mbox{45\arcsec}}}\right)^{-2} \frac{\cos i}{0.1
+ 0.9\left(\frac{L}{L_c}\right)^{0.3}} \mbox{ .}
\end{eqnarray}

\noindent Here $L_c = 6.4 \times 10^{21}$ W Hz$^{-1}$, the approximate
RC luminosity of an $L_*$ galaxy, and the correction applies only to
galaxies with radio luminosities below $L_c$. This correction is
obtained by comparing the RC to the infrared luminosity of a galaxy,
which is in turn calibrated against its UV emission. Because smaller
galaxies reprocess less of their UV light into infrared light, they
have more star formation per unit infrared luminosity, thus the
correction factor increases with decreasing galaxy luminosity (the
connection to the RC is via the empirical FIR-to-RC correlation). To
correct \xco\ by metallicity effects we employ the dependence found by
\citet[][]{WI95}

\begin{eqnarray}
\log \frac{X_{CO}}{X_{CO}^{MW}} = {5.95 - 0.67 \times [12 + \log
\mbox{(O/H)}}] \nonumber
\end{eqnarray}

\noindent where we have estimated metallicities from the absolute B
magnitudes of the dwarf galaxies using the relationship found by
\citet[][]{RM95} given in Equation \ref{RMZeq} above. We find that the
power law that best describes the relationship between the adjusted
$\Sigma_{SFR}$ and $\Sigma_{Mol}$ is

\begin{equation}
\Sigma_{SFR}^{\mathrm{Small}} = 10^{-3.6 \pm 0.3} \Sigma_{Mol}^{1.3
\pm 0.3}
\end{equation}

\noindent which agrees, within the uncertainties, with the uncorrected
results.

After applying reasonable corrections to Equations \ref{RCtoSFR} and
\ref{COtoMol}, the actual values of both $\Sigma_{Mol}$ and
$\Sigma_{SFR}$ have both increased by factors of $\sim 3$ for each
dwarf galaxy, so that we derive the same underlying relationship
between these two quantities. If we apply either correction without
the other to the dwarf galaxies, we get a difference in the
coefficients with no change in the exponent $b$ within the
uncertainties ($\Sigma_{SFR} = 10^{-4.0 \pm 0.3}\Sigma_{Mol}^{1.4 \pm
0.3}$ for only the \xco\ correction, and $\Sigma_{SFR} = 10^{-3.0 \pm
0.2}\Sigma_{Mol}^{1.1 \pm 0.4}$ for only the RC correction). Thus, if
only one correction holds, dwarf galaxies form a different mass of
stars per unit molecular gas than large galaxies --- less if only
\xco\ is a strong function of metallicity, more if only the SFR.

Therefore, we find that the power law index relating $\Sigma_{Mol}$ to
$\Sigma_{SFR}$ appears to be roughly the same, $\sim 1.1-1.4$,
irrespective of mild changes in the assumed \co-to-H$_2$ and RC-to-SFR
conversions. Plausible corrections to both the SFR and H$_2$ estimates
based on the RC and the CO yields a Schmidt Law coefficient that is
identical to that found with no corrections, implying that dwarf and
large galaxies show the same star formation surface density for a
given H$_2$ surface density.  This result breaks down if only one of
the SFR or the H$_2$ estimates are corrected. We do note that the
majority of dwarf galaxies considered here are dwarf spirals. An
extension of this analysis to dimmer dwarf irregular galaxies would
probe a yet more extreme set of physical conditions and be of
considerable interest, but this will require millimeter wave
telescopes with increased sensitivity.

% SUMMARY AND CONCLUSIONS
\section{Summary and Conclusions}

We report the detection of molecular gas in $28$ dwarf galaxies with
no previously published \co\ detections and the marginal detection of
$16$ more. We also present upper limits for $77$ dwarf galaxies that
we did not detect. These data increase by more than 50\% the number of
published \co\ detections of dwarf galaxies. Most detections are late
type spirals (Hubble type $\sim$Sc), of about the same stellar and
dynamical mass of the Large Magellanic Cloud, with considerably
brighter CO emission but somewhat lower star formation activity, and
located at a distance of $\sim11$ Mpc.  Nondetected galaxies tend to
be smaller than detections, with slightly later Hubble types (usually
Irr).

The most significant differences between the detections and the
nondetections appear to be the stellar mass, traced by $K$- and
$B$-band luminosities, and high mass star formation rate, traced by
their FIR emission and (less significantly) radio continuum
luminosities. The correlations we observe between the presence of \co\
emission and other galaxy parameters, such as Hubble type or mass to
light ratio can largely be ascribed to the fact that these properties
are also correlated with luminosity. The tight correlation between the
molecular gas mass and the $K$-band luminosity may result from the
dominant role played by the stellar disk in setting the midplane
pressure or equivalently local density of the atomic gas.  Thus, dwarf
galaxies tend to have roughly the same amount of molecular gas per
unit stellar mass as large spirals, although their ISMs are dominated
by large reservoirs of atomic gas and the molecular gas makes up only
a small fraction of the total gas mass.

Given the distances to our targets and the sensitivity of our survey,
we would not expect to detect galaxies with much less \co\ emission
than the LMC. Thus, we do not know if these correlations hold for
lower mass galaxies, but we see no evidence that we have ``hit a
wall'' in our attempt to detect \co\ in small galaxies. Both our
detections and the upper limits associated with our nondetections are
consistent with a nearly constant $M_{Mol}/L_{\odot,K} \approx 0.075$
throughout our sample.

Because the nondetections include the most ``primeval'' (metal poor,
low mass, and dynamically simple) galaxies in the sample, achieving
the order of magnitude increase in sensitivity necessary to (perhaps)
detect them in \co\ would be useful. If a change in \co\ properties
exists, it may well be at very low metallicities where CO ceases to
trace H$_2$. For instance, \citet*[][]{TKS98} suggested a sharp
increase in the CO-to-H$_2$ conversion factor, \xco\, below $Z\approx
1/10 Z_{\odot}$. Though we see a drop in detection rates with
decreasing metallicity (traced by B-magnitude) within our own sample,
the metallicity is covariant with many other galaxy properties and we
find no evidence for substantial changes in \xco between $Z_{\odot} >
Z > 1/4 Z_{\odot}$.

Combining our data with that of several previous \co\ surveys and the
NVSS, we find that dwarf galaxies with detected \co\ emission show the
same relationship between \co\ and the 1.4 GHz radio continuum as
large galaxies. This result suggests that there is a constant star
formation efficiency among dwarfs and large galaxies at a given
$\Sigma_{H_2}$. This conclusion is insensitive to the application of
small corrections to both the CO-to-H$_2$ and RC-to-SFR conversions,
although not to either conversion factor alone. Apparently, changes of
factors of $\sim5$ in metallicity and galaxy stellar mass are not
enough to markedly alter the $\Sigma_{SFR}$-to-$\Sigma_{Mol}$
relation.

\acknowledgements This research was partially supported by NSF grant
AST-0228963. We thank Aldo Apponi, Lucy Ziurys, and the team of ARO
12m operators headed by Paul Hart and including Sean Keel, Jon
Carlsen, John Downey and others. We thank the anonymous referee for
helpful feedback. This paper has made extensive use of three online
resources: (1) the NASA/IPAC Extragalactic Database (NED) which is
operated by the Jet Propulsion Laboratory, California Institute of
Technology, under contract with the National Aeronautics and Space
Administration; (2) the HyperLeda catalog, located on the World Wide
Web at http://www-obs.univ-lyon1.fr/hypercat/intro.html; and (3)
NASA's Astrophysics Data System (ADS). This publication makes use of
data products from the Two Micron All Sky Survey, which is a joint
project of the University of Massachusetts and the Infrared Processing
and Analysis Center/California Institute of Technology, funded by the
National Aeronautics and Space Administration and the National Science
Foundation. We would like to thank both the staff at the ARO 12m and
various members of the Berkeley Radio Astronomy Lab for numerous
helpful discussions and consultations. AL thanks Erik Rosolowksy, Ryan
Chornock, and Jason Wright for many helpful discussions of algorithms,
statistics, and science.

\clearpage 

\LongTables

\begin{center}
\begin{deluxetable}{l l l r r r r r r c}
\tablecaption{\label{dettable} Table of CO Observations}
\tabletypesize{\tiny}
\tablehead{ \colhead{Galaxy} & \colhead{RA} & 
\colhead{DEC} & \colhead{V$_{\mbox{LSR}}$} & 
\colhead{M$_{\mbox{B}}$} & \colhead{$\log$ L$_{\mbox{FIR}}$} &
\colhead{F$_{1.4}$} & \colhead{V$_{\mbox{rot}}$} & 
\colhead{I$_{\mbox{CO}}$} & \colhead{Detected} \\
& \colhead{(J2000)} & 
\colhead{(J2000)} & \colhead{(\kmpers)} & 
& \colhead{(L$_\odot$)} &
\colhead{(mJy)} & \colhead{(\kmpers)} & 
\colhead{(K \kmpers)}}
\startdata

 NGC 14 & 00 08 46.2 & 15 48 55.4 &  864.4 &  -17.9 &    8.4 &    2.8 &   51.4 & $\leq$   0.63 & N\\
 NGC 100 & 00 24 02.7 & 16 29 11.3 &  842.0 &  -17.9 &  $\ldots$  &  $< 1.0$  &   97.1 & $\leq$   0.44 & N\\
 UGC 1200 & 01 42 48.3 & 13 09 19.4 &  808.1 &  -16.2 &  $\ldots$  &    1.6 &   62.2 & $\leq$   0.81 & N\\
 IC 1727 & 01 47 29.9 & 27 19 59.1 &  346.4 &  -17.4 &    7.6 &    1.2 &   53.1 & $\leq$   0.60 & N\\
 UGC 1281 & 01 49 31.6 & 32 35 16.4 &  157.1 &  -15.8 &  $\ldots$  &    1.1 &   50.5 & $\leq$   0.80 & N\\
 NGC 784 & 02 01 16.7 & 28 50 14.2 &  197.6 &  -16.0 &    7.2 &    3.3 &   42.8 & $\leq$   0.76 & N\\
 NGC 949 & 02 30 48.8 & 37 08 12.1 &  611.1 &  -18.1 &    8.8 &   13.5 &  100.5 &   1.87 $\pm$   0.32 & Y\\
 NGC 959 & 02 32 23.8 & 35 29 41.6 &  601.7 &  -17.3 &    8.3 &    3.0 &   73.8 &   0.86 $\pm$   0.17 & Y\\
 UGC 2023 & 02 33 18.1 & 33 29 25.8 &  605.7 &  -16.4 &    7.8 &  $< 1.0$  &   46.1 & $\leq$   0.75 & N\\
 UGC 2082 & 02 36 16.3 & 25 25 24.2 &  706.7 &  -18.0 &    8.0 &    1.4 &   86.1 &   1.24 $\pm$   0.29 & M\\
 NGC 1012 & 02 39 14.9 & 30 09 05.0 &  978.3 &  -18.2 &    9.2 &   27.8 &   97.5 &   1.50 $\pm$   0.29 & Y\\
 NGC 1036 & 02 40 29.1 & 19 17 49.2 &  785.0 &  -17.1 &    8.5 &    4.3 &   83.8 &   0.87 $\pm$   0.17 & Y\\
 UGC 2259 & 02 47 55.5 & 37 32 17.5 &  584.9 &  -15.2 &  $\ldots$  &  $< 1.0$  &   92.5 & $\leq$   0.82 & N\\
 NGC 1156 & 02 59 42.2 & 25 14 15.3 &  373.6 &  -17.6 &    8.5 &   12.1 &   57.1 &   0.76 $\pm$   0.16 & Y\\
 NGC 1560 & 04 32 47.7 & 71 52 45.8 &  -36.9 &  -16.5 &  $\ldots$  &    3.3 &   76.2 &   1.36 $\pm$   0.40 & M\\
 UGC 3137 & 04 46 15.6 & 76 25 06.6 &  992.3 &  -17.2 &  $\ldots$  &  $< 1.0$  &   90.2 &   0.32 $\pm$   0.11 & M\\
 UGCA 105 & 05 14 14.8 & 62 34 48.3 &  111.5 &  -15.3 &  $\ldots$  &    2.1 &   51.6 &   0.61 $\pm$   0.16 & M\\
 UGC 3371 & 05 56 36.6 & 75 18 58.6 &  815.9 &  -16.9 &  $\ldots$  &  $< 1.0$  &   82.6 & $\leq$   0.58 & N\\
 UGCA 130 & 06 42 15.5 & 75 37 24.9 &  792.9 &  -15.6 &  $\ldots$  &  $< 1.0$  &   35.2 & $\leq$   1.00 & N\\
 NGC 2344 & 07 12 28.7 & 47 10 00.1 &  971.6 &  -18.3 &    8.4 &    1.8 &  153.8 &   1.24 $\pm$   0.23 & Y\\
 NGC 2366 & 07 28 51.9 & 69 12 30.9 &   99.6 &  -16.8 &    8.1 &    2.4 &   42.9 & $\leq$   1.35 & N\\
 NGC 2500 & 08 01 52.8 & 50 44 14.9 &  512.0 &  -17.8 &    8.6 &    3.6 &  127.4 & $\leq$   1.09 & N\\
 UGC 4278 & 08 13 59.0 & 45 44 37.6 &  559.2 &  -18.2 &    7.8 &  $< 1.0$  &   79.4 & $\leq$   0.56 & N\\
 NGC 2541 & 08 14 40.1 & 49 03 40.3 &  558.5 &  -18.2 &    8.5 &    1.6 &   93.1 & $\leq$   0.95 & N\\
 UGC 4305 & 08 19 03.9 & 70 43 08.7 &  157.9 &  -17.7 &    7.8 &    1.2 &   35.1 & $\leq$   1.32 & N\\
 NGC 2552 & 08 19 19.8 & 50 00 27.3 &  519.8 &  -17.5 &    8.0 &  $< 1.0$  &   62.7 &   0.36 $\pm$   0.10 & M\\
 UGC 4459 & 08 34 07.1 & 66 10 55.1 &   18.9 &  $\ldots$  &  $\ldots$  &    1.1 &  $\ldots$  & $\leq$   0.90 & N\\
 UGC 4499 & 08 37 41.4 & 51 39 09.7 &  691.7 &  -15.9 &    8.0 &    2.0 &   61.2 &   0.79 $\pm$   0.23 & M\\
 UGC 4514 & 08 39 37.7 & 53 27 24.1 &  693.5 &  -16.3 &    7.9 &    1.1 &   72.6 & $\leq$   0.88 & N\\
 UGC 5151 & 09 40 27.1 & 48 20 13.5 &  776.0 &  -17.0 &    8.2 &    2.1 &   84.1 & $\leq$   1.17 & N\\
 UGC 5272 & 09 50 22.4 & 31 29 16.0 &  520.1 &  -15.3 &    7.5 &    2.0 &   38.5 & $\leq$   0.75 & N\\
 UGC 5414 & 10 03 57.0 & 40 45 20.8 &  610.2 &  -16.2 &    7.9 &    1.8 &   53.8 & $\leq$   0.72 & N\\
 UGC 5456 & 10 07 19.7 & 10 21 44.2 &  540.4 &  -15.7 &    7.8 &    1.3 &   31.0 & $\leq$   1.00 & N\\
 NGC 3239 & 10 25 05.4 & 17 09 43.9 &  753.4 &  -19.0 &    8.9 &    8.6 &   79.3 & $\leq$   0.53 & N\\
 IC 2574 & 10 28 21.5 & 68 24 41.0 &   47.4 &  -17.6 &  $\ldots$  &  $< 1.0$  &   65.2 &   0.86 $\pm$   0.17 & Y\\
 DDO 82 & 10 30 34.5 & 70 37 13.4 &  180.0 &  $\ldots$  &  $\ldots$  &  $< 1.0$  &  $\ldots$  & $\leq$   0.96 & N\\
 NGC 3264 & 10 32 19.7 & 56 05 03.4 &  939.6 &  -18.2 &    8.3 &    2.1 &   53.3 & $\leq$   1.17 & N\\
 UGC 5829 & 10 42 42.1 & 34 26 56.0 &  624.7 &  -16.5 &    7.8 &    1.2 &   36.2 & $\leq$   1.20 & N\\
 NGC 3413 & 10 51 20.7 & 32 45 59.3 &  645.7 &  -17.7 &    8.3 &    4.7 &   72.3 & $\leq$   1.09 & N\\
 UGC 5986 & 10 52 30.9 & 36 37 08.7 &  615.5 &  -19.3 &    9.1 &   22.2 &  118.9 & $\leq$   0.81 & N\\
 NGC 3510 & 11 03 43.5 & 28 53 07.0 &  704.3 &  -16.7 &    8.2 &    3.1 &   83.2 & $\leq$   0.79 & N\\
 UGCA 225 & 11 04 58.2 & 29 08 17.1 &  645.8 &  -14.6 &  $\ldots$  &    2.0 &   38.5 & $\leq$   1.24 & N\\
 UGC 6446 & 11 26 40.4 & 53 44 48.4 &  644.3 &  -16.8 &  $\ldots$  &  $< 1.0$  &   80.2 & $\leq$   0.54 & N\\
 UGC 6448 & 11 26 50.3 & 64 08 19.6 &  989.6 &  -16.3 &    8.2 &  $< 1.0$  &   38.5 & $\leq$   0.65 & N\\
 UGC 6456 & 11 27 58.8 & 78 59 38.4 &  -93.4 &  -11.2 &  $\ldots$  &    1.2 &   20.7 & $\leq$   0.99 & N\\
 NGC 3773 & 11 38 12.9 & 12 06 43.5 &  975.6 &  -16.9 &    8.6 &    5.6 &   74.3 & $\leq$   1.01 & N\\
 NGC 3782 & 11 39 20.5 & 46 30 51.4 &  737.1 &  -17.6 &    8.6 &    4.6 &   64.9 &   1.19 $\pm$   0.33 & M\\
 UGC 6628 & 11 40 05.7 & 45 56 32.6 &  849.8 &  -17.6 &    8.1 &  $< 1.0$  &   25.6 & $\leq$   0.60 & N\\
 NGC 3870 & 11 45 56.6 & 50 11 57.4 &  755.0 &  -17.3 &    8.6 &    4.8 &   67.6 &   0.52 $\pm$   0.11 & Y\\
 NGC 3906 & 11 49 40.9 & 48 25 34.3 &  950.9 &  -17.3 &    8.4 &    2.0 &  127.6 &   0.59 $\pm$   0.18 & M\\
 NGC 3913 & 11 50 38.9 & 55 21 13.6 &  954.0 &  -17.8 &    8.4 &  $< 1.0$  &   34.8 &   1.98 $\pm$   0.24 & Y\\
 UGC 6900 & 11 55 39.0 & 31 31 07.6 &  589.6 &  -15.5 &  $\ldots$  &  $< 1.0$  &   48.5 & $\leq$   0.81 & N\\
 UGC 6917 & 11 56 27.7 & 50 25 43.6 &  911.2 &  -18.1 &    8.2 &    1.1 &   91.8 & $\leq$   1.29 & N\\
 NGC 3985 & 11 56 42.0 & 48 20 01.6 &  947.3 &  -18.1 &    8.8 &    6.5 &   84.3 &   0.88 $\pm$   0.19 & Y\\
 NGC 3990 & 11 57 35.6 & 55 27 29.1 &  695.0 &  $\ldots$  &  $\ldots$  &  $< 1.0$  &  $\ldots$  & $\leq$   1.12 & N\\
 NGC 4068 & 12 04 01.0 & 52 35 16.4 &  210.2 &  -15.0 &    7.3 &  $< 1.0$  &   27.1 &   0.68 $\pm$   0.18 & M\\
 NGC 4080 & 12 04 51.8 & 26 59 33.3 &  585.7 &  -16.2 &    7.8 &    1.7 &   82.2 &   1.88 $\pm$   0.43 & Y\\
 NGC 4136 & 12 09 17.7 & 29 55 41.1 &  607.7 &  -17.5 &    8.5 &  $< 1.0$  &   93.3 &   1.59 $\pm$   0.25 & Y\\
 NGC 4144 & 12 09 58.6 & 46 27 27.0 &  267.1 &  -17.6 &    8.1 &    4.9 &   76.0 &   1.06 $\pm$   0.22 & Y\\
 NGC 4150 & 12 10 33.6 & 30 24 05.7 &  226.0 &  $\ldots$  &    7.8 &    1.4 &  $\ldots$  &   2.38 $\pm$   0.42 & Y\\
 NGC 4190 & 12 13 44.1 & 36 37 53.7 &  230.0 &  -15.0 &    7.4 &    4.6 &   50.5 & $\leq$   1.28 & N\\
 UGC 7261 & 12 15 14.3 & 20 39 32.0 &  838.0 &  -16.8 &    8.4 &    1.0 &   68.4 & $\leq$   0.63 & N\\
 NGC 4218 & 12 15 46.1 & 48 07 54.1 &  724.8 &  -16.9 &    8.5 &    4.9 &   69.2 & $\leq$   0.95 & N\\
 NGC 4299 & 12 21 40.4 & 11 30 10.8 &  231.1 &  -15.4 &  $\ldots$  &   10.1 &  122.7 & $\leq$   0.77 & N\\
 NGC 4309 & 12 22 12.3 & 07 08 39.1 &  871.7 &  -16.6 &    8.4 &    2.3 &  109.4 &   3.01 $\pm$   0.54 & Y\\
 NGC 4310 & 12 22 26.3 & 29 12 29.8 &  887.1 &  -17.4 &    8.6 &    2.2 &   78.9 &   3.31 $\pm$   0.52 & Y\\
 UGC 7490 & 12 24 24.7 & 70 20 02.0 &  466.7 &  -16.4 &    7.8 &  $< 1.0$  &   36.1 & $\leq$   1.10 & N\\
 NGC 4395 & 12 25 49.0 & 33 32 49.2 &  319.7 &  -18.0 &  $\ldots$  &    2.1 &   57.4 & $\leq$   0.93 & N\\
 NGC 4396 & 12 25 59.1 & 15 40 15.5 & -124.6 &  $\ldots$  &    5.3 &    8.8 &   81.8 &   2.37 $\pm$   0.48 & Y\\
 UGC 7559 & 12 27 04.7 & 37 08 38.7 &  218.1 &  -14.7 &  $\ldots$  &  $< 1.0$  &   30.1 & $\leq$   0.98 & N\\
 UGC 7557 & 12 27 11.3 & 07 15 44.9 &  932.9 &  -17.8 &    8.0 &    1.9 &   63.8 & $\leq$   0.87 & N\\
 UGC 7599 & 12 28 28.1 & 37 14 00.6 &  277.8 &  -14.2 &  $\ldots$  &  $< 1.0$  &   29.8 & $\leq$   1.27 & N\\
 UGC 7603 & 12 28 44.0 & 22 49 17.0 &  642.1 &  -16.7 &    8.0 &    1.9 &   51.2 & $\leq$   0.69 & N\\
 IC 3414 & 12 29 29.0 & 06 46 16.6 &  537.1 &  -15.6 &    7.5 &  $< 1.0$  &   52.7 & $\leq$   2.27 & N\\
 UGC 7690 & 12 32 26.8 & 42 42 18.0 &  537.4 &  -16.9 &    8.0 &    1.6 &   71.8 & $\leq$   0.99 & N\\
 UGC 7698 & 12 32 54.4 & 31 32 30.8 &  332.7 &  -16.2 &  $\ldots$  &    1.1 &   29.1 & $\leq$   0.80 & N\\
 NGC 4509 & 12 33 06.7 & 32 05 27.9 &  935.1 &  -15.8 &    8.3 &    2.8 &   40.9 & $\leq$   1.25 & N\\
 NGC 4534 & 12 34 05.4 & 35 31 05.1 &  801.9 &  -17.8 &    8.4 &    2.0 &   74.8 & $\leq$   1.40 & N\\
 NGC 4618 & 12 41 32.5 & 41 08 57.1 &  543.1 &  -18.8 &    8.9 &    6.6 &   67.6 &   1.23 $\pm$   0.21 & Y\\
 NGC 4625 & 12 41 52.6 & 41 16 27.1 &  610.6 &  -17.3 &    8.5 &    3.5 &   47.4 &   2.26 $\pm$   0.26 & Y\\
 NGC 4630 & 12 42 31.1 & 03 57 33.1 &  738.7 &  -17.1 &    8.7 &    8.2 &   75.9 &   4.45 $\pm$   0.38 & Y\\
 NGC 4633 & 12 42 37.0 & 14 21 19.8 &  290.3 &  -15.4 &  $\ldots$  &    2.0 &   91.4 & $\leq$   0.94 & N\\
 NGC 4635 & 12 42 39.2 & 19 56 43.0 &  960.0 &  -17.9 &    8.4 &    1.1 &  101.4 &   1.71 $\pm$   0.29 & Y\\
 UGCA 294 & 12 44 38.1 & 28 28 21.3 &  944.9 &  -15.4 &  $\ldots$  &  $< 1.0$  &   42.1 & $\leq$   0.59 & N\\
 IC 3742 & 12 45 31.7 & 13 19 51.9 &  964.7 &  -17.2 &    8.1 &  $< 1.0$  &   81.3 & $\leq$   0.78 & N\\
 UGCA 298 & 12 46 55.3 & 26 33 51.4 &  830.2 &  $\ldots$  &  $\ldots$  &  $< 1.0$  &  $\ldots$  & $\leq$   1.07 & N\\
 NGC 4688 & 12 47 46.5 & 04 20 08.1 &  986.3 &  -17.4 &    8.6 &    1.9 &   60.1 & $\leq$   1.20 & N\\
 NGC 4701 & 12 49 11.6 & 03 23 18.9 &  723.1 &  -17.5 &    8.8 &   11.9 &  108.4 &   4.85 $\pm$   0.56 & Y\\
 NGC 4713 & 12 49 57.6 & 05 18 41.4 &  652.2 &  -17.9 &    8.9 &   18.5 &  119.4 &   3.61 $\pm$   0.39 & Y\\
 NGC 4765 & 12 53 14.6 & 04 27 46.7 &  724.5 &  -17.0 &    8.6 &   14.4 &   51.4 &   0.85 $\pm$   0.26 & M\\
 UGC 8146 & 13 02 07.8 & 58 42 00.0 &  668.8 &  -16.7 &  $\ldots$  &  $< 1.0$  &   77.1 & $\leq$   1.27 & N\\
 UGC 8246 & 13 10 04.3 & 34 10 49.4 &  799.5 &  -16.4 &    7.9 &  $< 1.0$  &   62.7 & $\leq$   1.05 & N\\
 IC 4213 & 13 12 11.1 & 35 40 18.1 &  814.8 &  -17.6 &    8.0 &    1.0 &   81.7 &   0.84 $\pm$   0.20 & Y\\
 DDO 166 & 13 13 17.8 & 36 12 56.5 &  945.5 &  -17.3 &    8.3 &  $< 1.0$  &   52.1 &   2.01 $\pm$   0.48 & M\\
 UGC 8320 & 13 14 27.8 & 45 55 11.2 &  194.4 &  -16.2 &  $\ldots$  &    1.3 &   50.1 & $\leq$   0.97 & N\\
 UGC 8323 & 13 14 48.4 & 34 52 51.6 &  858.3 &  -16.3 &    8.2 &    2.1 &   50.2 & $\leq$   1.54 & N\\
 NGC 5107 & 13 21 24.5 & 38 32 18.6 &  945.6 &  -17.2 &    8.6 &    3.6 &   67.8 & $\leq$   0.87 & N\\
 UGC 8490 & 13 29 36.4 & 58 25 12.7 &  202.9 &  -17.5 &    8.3 &    3.6 &   58.3 & $\leq$   0.78 & N\\
 UGC 8507 & 13 30 58.6 & 19 26 11.4 &  999.2 &  -17.3 &    8.1 &    1.0 &   44.8 & $\leq$   1.08 & N\\
 NGC 5238 & 13 34 42.5 & 51 36 48.5 &  233.0 &  -14.9 &  $\ldots$  &  $< 1.0$  &   21.6 & $\leq$   1.01 & N\\
 NGC 5338 & 13 53 26.5 & 05 12 27.3 &  815.8 &  -16.6 &    8.0 &    1.1 &   32.7 &   0.97 $\pm$   0.18 & Y\\
 NGC 5474 & 14 05 01.5 & 53 39 44.2 &  276.3 &  -17.6 &    8.2 &    1.8 &   22.1 & $\leq$   0.40 & N\\
 NGC 5477 & 14 05 33.1 & 54 27 40.3 &  305.0 &  -14.9 &    7.5 &  $< 1.0$  &   30.9 & $\leq$   0.97 & N\\
 NGC 5585 & 14 19 47.6 & 56 43 44.7 &  304.9 &  -18.2 &    8.2 &    2.5 &   96.8 & $\leq$   0.99 & N\\
 UGC 9405 & 14 35 24.5 & 57 15 16.5 &  221.1 &  $\ldots$  &  $\ldots$  &  $< 1.0$  &   36.2 & $\leq$   0.57 & N\\
 NGC 5832 & 14 57 45.7 & 71 40 53.0 &  448.9 &  -16.3 &    8.0 &    2.7 &   74.5 &   1.20 $\pm$   0.29 & M\\
 NGC 5949 & 15 28 00.6 & 64 45 47.8 &  429.8 &  -17.1 &    8.4 &    3.4 &   88.3 &   1.86 $\pm$   0.51 & M\\
 NGC 5963 & 15 33 27.7 & 56 33 32.7 &  655.1 &  -17.8 &    8.8 &    6.5 &  121.1 &   2.29 $\pm$   0.30 & Y\\
 UGCA 410 & 15 37 04.1 & 55 15 48.6 &  660.8 &  -15.3 &  $\ldots$  &  $< 1.0$  &   61.5 & $\leq$   0.93 & N\\
 UGC 10310 & 16 16 18.1 & 47 02 44.5 &  713.8 &  -17.1 &    8.0 &  $< 1.0$  &   55.8 & $\leq$   0.76 & N\\
 UGC 10445 & 16 33 47.7 & 28 59 05.2 &  961.8 &  -17.3 &    8.4 &  $< 1.0$  &   82.4 &   0.60 $\pm$   0.17 & M\\
 NGC 6255 & 16 54 47.3 & 36 30 05.3 &  915.3 &  -17.8 &    8.4 &    1.2 &   98.6 & $\leq$   1.13 & N\\
 UGC 10806 & 17 18 51.3 & 49 53 02.3 &  929.5 &  -17.4 &    8.4 &    2.3 &   71.1 & $\leq$   0.75 & N\\
 NGC 6689 & 18 34 49.8 & 70 31 28.9 &  489.2 &  -17.9 &    8.3 &    1.8 &   91.0 &   1.49 $\pm$   0.28 & Y\\
 NGC 6789 & 19 16 41.6 & 63 58 22.4 & -148.5 &  $\ldots$  &  $\ldots$  &  $< 1.0$  &  $\ldots$  &   1.04 $\pm$ 0.27 & M\\
 UGC 11891 & 22 03 33.7 & 43 44 56.7 &  461.4 &  -16.1 &  $\ldots$  &    1.2 &   86.9 &   0.33 $\pm$   0.11 & M\\
 NGC 7292 & 22 28 25.9 & 30 17 28.6 &  985.6 &  -18.1 &    8.7 &    5.0 &   44.2 & $\leq$   0.77 & N\\
 UGC 12060 & 22 30 33.6 & 33 49 17.0 &  884.0 &  -15.5 &  $\ldots$  &  $< 1.0$  &   52.5 &   1.12 $\pm$   0.18 & Y\\
 DDO 213 & 22 34 10.8 & 32 51 41.0 &  804.7 &  -16.9 &  $\ldots$  &  $< 1.0$  &   52.4 & $\leq$   0.81 & N\\
 UGC 12632 & 23 29 58.7 & 40 59 25.4 &  422.2 &  -17.4 &  $\ldots$  &  $< 1.0$  &   79.3 & $\leq$   0.71 & N\\
 UGC 12732 & 23 40 39.9 & 26 14 10.3 &  748.5 &  -16.5 &  $\ldots$  &  $< 1.0$  &   95.1 & $\leq$   0.68 & N\\

\enddata
\end{deluxetable}
\end{center}

\clearpage
\begin{landscape}
\begin{deluxetable}{l c c c c c c c}
\tabletypesize{\tiny}
%\rotate
\tablewidth{0pt}
\tablecaption{\label{CorrTab} Rank Correlations Between \co\ and
other Galaxy Properties}
\tablehead{ \colhead{Property} & \colhead{with $M_{Mol}$} &
\colhead{with $M_{Mol}/L_{K}$} & \colhead{with $M_{Mol}/L_{B}$} &
\colhead{with $M_{Mol}/M_{dyn}$} & \colhead{with $M_{Mol}/M_{HI}$} &
\colhead{with $M_{Mol}/L_{RC}$} & \colhead{with $M_{Mol}/L_{FIR}$} \\ \\
& \colhead{Molecular Gas} & \colhead{Molecular Gas} & & 
\colhead{Molecular Gas} & \colhead{Molecular
Gas} & \colhead{Molecular Gas} & \\
& \colhead{Mass} & \colhead{per unit} & & \colhead{per unit}
& \colhead{to Atomic Gas} & \colhead{Depletion Time} & \\
& & \colhead{Stellar Mass} & & \colhead{Dynamical Mass}
& \colhead{Ratio} & & }

\startdata
Biases & & & & & \\
\hline
Virgocentric Velocity/Distance & $\textbf{+0.62}$\tablenotemark{a} & $+0.37$ & $+0.38$ & $+0.41$ & $+0.32$ & \nodata & \nodata \\
Angular Size & \nodata & $-0.42$ & $\textbf{-0.35}$\tablenotemark{a} & $-0.39$ & \nodata & \nodata & \nodata \\
& & & & & \\
\hline
Global Quantities & & & & & \\
\hline
Hubble Type & $-0.49$ & \nodata & $-0.26$ & \nodata & $\textbf{-0.55}$\tablenotemark{a} & \nodata & \nodata \\
Dynamical Mass & $+0.69$ & \nodata & \nodata & \nodata & $+0.36$ & \nodata & $+0.30$ \\
K-band Luminosity & $\textbf{+0.85}$\tablenotemark{a} & \nodata & $+0.26$ & \nodata & $+0.50$ & \nodata & $+0.30$ \\
B-band Luminosity & $\textbf{+0.79}$\tablenotemark{a} & \nodata & \nodata & \nodata & $+0.41$ & \nodata & $+0.29$ \\
Linear Diameter & $\textbf{+0.68}$\tablenotemark{a} & \nodata & \nodata & \nodata & $+0.22$ & \nodata & $+0.26$ \\
FIR Luminosity & $\textbf{+0.90}$\tablenotemark{a} & $+0.47$ & $+0.57$ & $+0.55$ & $+0.60$ & $-0.42$ & \nodata \\
RC Luminosity & $\textbf{+0.86}$\tablenotemark{a} & $+0.35$ & $+0.45$ & $+0.43$ & $+0.53$ & $-0.48$ & \nodata \\
HI Luminosity & $+0.55$ & \nodata & \nodata & \nodata & $\textbf{(-0.12)}$\tablenotemark{a,b} & \nodata & \nodata \\
& & & & & \\
\hline
Colors & & & & & \\
\hline
$B - V$ & $+0.35$ & \nodata & \nodata & \nodata & $+0.53$ & \nodata & $+0.35$ \\
$B - K$ & $+0.46$ & \nodata & $\textbf{+0.56}$\tablenotemark{a} & $+0.31$ & $+0.57$ & \nodata & \nodata \\
FIR (60 $\mu$m) / FIR (100 $\mu$m) & \nodata & $+0.42$ & $+0.40$ & $+0.46$ & $+0.31$ & $-0.55$ & $-0.35$ \\
$M_{HI}/L_B$ & $-0.42$ & \nodata & \nodata & \nodata & \nodata & \nodata & $-0.24$ \\
$M_{HI}/L_K$ & $-0.50$ & \nodata & $\textbf{-0.30}$\tablenotemark{a} & \nodata & \nodata & \nodata & $-0.25$ \\
$M_{dyn}/L_B$ & \nodata & \nodata & \nodata & \nodata & \nodata & \nodata & \nodata \\
$M_{dyn}/L_K$ & $\textbf{-0.28}$\tablenotemark{a} & \nodata & $-0.27$ & \nodata & $-0.32$ & $+0.23$ & \nodata \\
\hline
Median Values & $M_{\odot}$ & $M_{\odot}/L_{K,\odot}$ &
$M_{\odot}/L_{B,\odot}$ & $M_{\odot}/M_{\odot} $ &
$M_{\odot}/M_{\odot}$ & $M_{\odot}/L_{\nu,1.4} (10^{12} \mathrm{W~Hz}^{-1})$ & $M_{\odot}/L_{\odot} $\\
\hline
Dwarfs & $3 \pm 0.5 \times 10^8$ & $0.065 \pm 0.008$ & $0.13 \pm 0.02$
& $0.037 \pm 0.007$ & $0.30 \pm 0.05$ & $1.4 \pm 0.2$ & $0.47 \pm 0.08$ \\
Large Spirals & $5 \pm 0.5 \times 10^9$ & $0.075 \pm 0.005$ & $0.16
\pm 0.01$ & $0.040 \pm 0.003$ & $1.5 \pm 0.1$ & $1.0 \pm 0.05$ & $0.64 \pm 0.03$ \\
\enddata

\tablecomments{This table shows rank correlation coefficients for a
combined sample of large spirals and dwarf galaxies. Typical $1\sigma$
uncertainties for the values in this table are $\pm 0.07$. Only Values
significant at $> 3\sigma$ level are shown.}

\tablenotetext{a}{{\bf Boldfaced} values are significant at the
$3\sigma$ level or higher {\it in the sample of dwarfs alone}. Typical
$1\sigma$ uncertainties for dwarf galaxies alone are $\pm 0.16$, so
these values have rank correlation coefficients of $\gtrsim 0.5$ among
the dwarfs.}

\tablenotetext{b}{$M_{HI}$ and $M_{Mol}/M_{HI}$ are correlated at $>
3\sigma$ significance within the subset of dwarfs but not within the
larger sample. This is the only pair of properties for which we find
this result.}

\end{deluxetable}
\clearpage
\end{landscape}

\begin{deluxetable}{l c c c c}
\tabletypesize{\small}
\tablewidth{0pt}
\tablecolumns{5}
\tablecaption{\label{SlopeTab} Power Law Relations Between \co\ and
other Galaxy Properties}

\tablehead{ \multicolumn{1}{c}{Property} &
\multicolumn{2}{c}{Exponent} & \multicolumn{2}{c}{Scatter About Best
Fit (dex)} \\
\multicolumn{1}{c}{} &
\multicolumn{2}{c}{$\overbrace{\phm{SpanningSpann}}$} &
\multicolumn{2}{c}{$\overbrace{\phm{SpanningSpann}}$} \\
\multicolumn{1}{c}{} &
\multicolumn{1}{c}{All Galaxies} & \multicolumn{1}{c}{Dwarfs} &
\multicolumn{1}{c}{All Galaxies} & \multicolumn{1}{c}{Dwarfs} }

\startdata

Dynamical Mass & $1.2 \pm 0.2$ & \nodata \tablenotemark{a} & 0.6 & 0.6\tablenotemark{a} \\
K-band Luminosity & $1.2 \pm 0.1$ & $1.3 \pm 0.2$ & 0.4 & 0.4 \\
B-band Luminosity & $1.2 \pm 0.3$ & $1.3 \pm 0.2$ & 0.5 & 0.5 \\
FIR Luminosity & $1.1 \pm 0.1$ & $1.0 \pm 0.3$ & 0.4 & 0.5 \\
RC Luminosity & $0.8 \pm 0.1$ & $0.9 \pm 1.6$ & 0.4 & 0.6 \\

\enddata

\tablenotetext{a}{The slope of the $M_{Mol}$ to $M_{dyn}$ relation is
very poorly determined in dwarfs. We quote the scatter among the
dwarfs using the fit to the combined sample.}

\end{deluxetable}

\begin{deluxetable}{l c c }
\tablecaption{\label{MedTab} Median Detection/Nondetection Properties}
\tablehead{ \colhead{Property} & \colhead{Median Detection Value} & 
\colhead{Median Nondetection Value}}
\startdata
Biases & & \\
\hline
Virgocentric Velocity Distance & $11.3 \pm 0.6$ Mpc & $10.9 \pm 0.5$ Mpc \\
RMS Noise & $7.7 \pm 0.4$ mK & $8.3 \pm 0.2$ mK \\
Angular Size & $2.4' \pm 0.8$ & $2.7' \pm 0.2$ \\
& & \\
\hline
Global Quantities & & \\
\hline
Hubble Type & $\sim$ Sc & Sm/Irr \\
Dynamical Mass ($\frac{V^2 R}{G}$) & $8 \pm 1 \times 10^9$ M$_{\odot}$
& $3 \pm 1 \times 10^9$ M$_{\odot}$ \\
Absolute K Magnitude & $-20.4 \pm 0.2$ & $-19.5 \pm 0.2$ \\
Absolute B Magnitude & $-17.6 \pm 0.1$ & $-16.8 \pm 0.2$ \\
Linear Diameter ($d_{25}$ & $8.5 \pm 0.8$ kpc & $7.7 \pm 0.7$ kpc \\
$L_{FIR}$ & $3.2 \pm 0.6 \times 10^{8} L_{\odot}$ & $1.5 \pm 0.2 \times 10^{8} L_{\odot}$ \\ 
$L_{RC}$ & $13 \pm 2 \times 10^{19}$ W Hz$^{-1}$ & $5 \pm 1 \times 10^{19}$ W Hz$^{-1}$ \\
$M_{HI}$ & $9 \pm 2 \times 10^8$ M$_{\odot}$ & $8 \pm 1 \times 10^8$ M$_{\odot}$ \\
& & \\
\hline
Colors & & \\
\hline
$B - V$ & $0.48 \pm 0.03$ & $0.43 \pm 0.02$ \\ 
$B - K$ & $2.8 \pm 0.1$ & $2.1 \pm 0.1$ \\
FIR (60 $\mu$m) / FIR (100 $\mu$m) & $0.40 \pm 0.02$ & $0.41 \pm 0.02$ \\
$M_{HI}/L_B$ & $0.40 \pm 0.06$ M$_{\odot}$/$L_{B\odot}$ & $0.87 \pm 0.09$ M$_{\odot}$/$L_{B\odot}$ \\
$M_{HI}/L_K$ & $0.24 \pm 0.04$ M$_{\odot}$/$L_{K\odot}$ & $0.77 \pm 0.18$ M$_{\odot}$/$L_{K\odot}$\\
$M_{dyn}/L_B$ & $4.5 \pm 0.8$ & $3.8 \pm 0.3$ \\
$M_{dyn}/L_K$ & $2.2 \pm 0.3$ & $4.2 \pm 0.8$ \\
& & \\
\hline
CO Properties & & \\
\hline
L$_{CO}$ & $1.9 \pm 0.6 \times 10^7$ K km s$^{-1}$ pc$^2$ & $< 6.9 \times 10^6$ K km s$^{-1}$ pc$^2$ ($2\sigma$) \\
M$_{Mol}$ & $8.2 \pm 2.6 \times 10^7$ M$_{\odot}$ & $< 3.0 \times
10^7$ M$_{\odot}$
\mbox{ } (2$\sigma$) \\
Molecular Gas Surface Density & $3.1 \pm 0.6$ M$_{\odot}$ pc$^{-2}$ & $< 1.6$ M$_{\odot}$ pc$^{-2}$ ($2\sigma$) \\
\enddata
\end{deluxetable}

\begin{deluxetable}{l c c}
\tablecaption{\label{DiffTab} Differences Between Detections and Nondetections}
\tablehead{ \colhead{Property} & \colhead{KS Significance} & 
\colhead{Student's t Significance}}
\startdata
Biases & & \\
\hline
Virgocentric Velocity/Distance & 0.35 & 0.67 \\
Angular Size & 0.68 & 0.88 \\
RMS Noise in CO Spectrum & 0.69 & 0.35 \\
& & \\
\hline
Global Quantities & & \\
\hline
Hubble Type & 0.01 & 0.01 \\
Dynamical Mass & 0.03 & 0.02 \\
Absolute K Magnitude & $\sim 10^{-4}$ & $\sim 10^{-6}$ \\
Absolute B Magnitude & 0.001 & 0.001 \\
Linear Diameter & 0.44 & 0.88 \\
$L_{FIR}$ & $\sim 10^{-4}$ & $\sim 10^{-4}$ \\
$L_{RC}$ & 0.03 & 0.03 \\
$M_{HI}$ & 0.58 & 0.46 \\
& & \\
\hline
Colors & & \\
\hline
$B - V$ & 0.34 & 0.14 \\ 
$B - K$ & 0.001 & $\sim 10^{-4}$ \\
FIR (60 $\mu$m) / FIR (100 $\mu$m) & 0.55 & 0.28 \\
$M_{HI}/L_B$ & $\sim 10^{-4}$ & $\sim 10^{-6}$ \\
$M_{HI}/L_K$ &  $\sim 10^{-4}$ & $\sim 10^{-5}$ \\
$M_{dyn}/L_B$ & 0.31 & 0.50 \\
$M_{dyn}/L_K$ & 0.05 & 0.03 \\
\enddata
\end{deluxetable}

\end{document}